\RequirePackage{fix-cm}
\documentclass[twocolumn]{svjour3}       
\smartqed  
\usepackage{graphicx, hyperref, breakurl}
\usepackage{epstopdf, bbm}
\usepackage{natbib, url ,amsmath, amssymb, latexsym, appendix}
\usepackage{float, multirow, subcaption} 
\floatstyle{plain}
\newfloat{Algorithm}{thp}{lop}
\floatname{Algorithm}{Algorithm}

\DeclareMathOperator*{\argmax}{arg\,max}
\newcommand{\ra}[1]{\renewcommand{\arraystretch}{#1}}
\allowdisplaybreaks
\begin{document}
\sloppy

\title{Stochastic variational inference for large-scale discrete choice models using adaptive batch sizes}


\author{Linda S. L. Tan}


\institute{Linda S. L. Tan \at
              Department of Statistics and Applied Probability \\
              National University of Singapore \\
                            Tel.: +65-6516-4416\\ 
                            Fax: +65-6872-3919\\  
              \email{statsll@nus.edu.sg}           
}

\date{Received: date / Accepted: date}

\maketitle

\begin{abstract} 
Discrete choice models describe the choices made by decision makers among alternatives and play an important role in transportation planning, marketing research and other applications. The mixed multinomial logit (MMNL) model is a popular discrete choice model that captures heterogeneity in the preferences of decision makers through random coefficients. While Markov chain Monte Carlo methods provide the Bayesian analogue to classical procedures for estimating MMNL models, computations can be prohibitively expensive for large datasets. Approximate inference can be obtained using variational methods at a lower computational cost with competitive accuracy. In this paper, we develop variational methods for estimating MMNL models that allow random coefficients to be correlated in the posterior and can be extended easily to large-scale datasets. We explore three alternatives: (1) Laplace variational inference, (2) nonconjugate variational message passing and (3) stochastic linear regression. Their performances are compared using real and simulated data. To accelerate convergence for large datasets, we develop stochastic variational inference for MMNL models using each of the above alternatives. Stochastic variational inference allows data to be processed in minibatches by optimizing global variational parameters using stochastic gradient approximation. A novel strategy for increasing minibatch sizes adaptively within stochastic variational inference is proposed.

\end{abstract}

\section{Introduction}
Discrete choice models form the basis for understanding the behavioural process that results in a choice made by a decision maker (or agent) among a finite set of alternatives. They are highly flexible and can be applied in a wide variety of choice situations. For example, the agents can be consumers choosing between different brands in a product category; or households selecting among different types of heating systems. Discrete choice models are widely used to predict demand for new systems in transportation planning \citep{Ben-Akiva1985}, develop pricing policies in marketing research \citep{McFadden1980}, elicit preferences for healthcare products and programmes in health economics \citep{Lancsar2008} and in many other applications.

The mixed multinomial logit (MMNL) model is a popular discrete choice model that captures heterogeneity in the preferences of decision makers through random coefficients \citep[see, e.g.][]{Train2009}. Choice probabilities of the MMNL model are expressed in the form 
\begin{equation*}
\int \frac{\exp(x_{hj}^T \beta_h)}{\sum_{j'=1}^J \exp(x_{hj'}^T \beta_h)}\, G(\beta_h) \;d\beta_h,
\end{equation*}
where $J$ denotes the number of alternatives, $x_{hj}$ is a $K \times 1$ vector of observed variables relating to alternative $j$ and agent $h$, and $\beta_h$ is a $K \times 1$ vector of random coefficients that vary over agents in the population with mixing distribution $G(\beta_h)$. The distribution $G$ may be discrete or continuous (e.g. normal, lognormal, triangular and uniform). The parameters of $G$, and in some cases, the values of $\beta_h$ (which represent the preferences of individual decision makers) are of interest. The importance of the MMNL model and its ability to accommodate heterogeneity is well established \citep[e.g.][]{Bhat1998, Brownstone1999}. \cite{McFadden2000} showed that any random utility model can be approximated to an arbitrary degree of accuracy by a MMNL model with appropriate choice of variables and mixing distribution.

The classical approach to estimating MMNL models is via maximization of the simulated likelihood function \cite[see, e.g.][]{McFadden2000}. This procedure can be difficult numerically; the algorithm may not converge due to various reasons and there is a risk of getting trapped in local maxima. Bayesian procedures avoid some of these issues, and consistency and efficiency in estimation can be attained under fewer restrictions \cite[see][Chap. 12]{Train2009}. Markov chain Monte Carlo (MCMC) methods provide the Bayesian analogue to classical procedures for estimating MMNL models. In the hierarchical Bayesian approach, draws from the posterior distribution can be obtained using Gibbs sampling and the Metropolis-Hastings algorithm \cite[see, e.g.][]{Rossi2005}. However, computations can be prohibitively expensive for large datasets, which are increasingly common. An alternative is to obtain approximate Bayesian inference via variational approximation methods \citep{Jordan1999}. \cite{Braun2010} showed that predictive inference for MMNL models can be obtained using variational methods at a lower computational cost but with accuracy close to that of MCMC. 

In this paper, we develop variational methods for estimating MMNL models that allow the coefficients in $\beta_h$ to be correlated in the posterior. Previously, \cite{Braun2010} considered MMNL models where $\beta_h$ is normally distributed and derived a variational Bayes \citep{Attias1999} procedure for posterior approximation. The approximating density was assumed to be of a factorized parametric form and the reverse Kullback-Leibler divergence between the true posterior and the variational approximation was minimized. This is equivalent to maximizing a lower bound on the log marginal likelihood. The intractable lower bound was approximated using the multivariate delta method for moments \citep{Bickel2007}. Maximization of the lower bound over individual-level variational parameters was performed using standard unconstrained convex optimization techniques, and the covariance matrix of $\beta_h$ was assumed to be diagonal in the variational posterior. This is due likely to the high computational cost of optimizing a full covariance matrix. 

We explore three alternatives that allow the posterior independence assumption among random coefficients to be relaxed at low computational cost: (1) Laplace variational inference, (2) nonconjugate variational message passing and (3) stochastic linear regression. The performances of these approaches are compared using real and simulated datasets. Laplace variational inference was first considered by \cite{Waterhouse1996}. \cite{Wang2013} formalized the approach for a class of nonconjugate models. This method uses Laplace approximations within the optimal density update in variational Bayes, and results in a Gaussian approximation of the posterior. Nonconjugate variational message passing is an algorithm proposed by \cite{Knowles2011} to extend variational Bayes to nonconjugate models. The variational posterior is assumed to be some member of the exponential family and variational parameters can be obtained using fast fixed point updates. We continue to use the delta method to approximate the intractable lower bound when using nonconjugate variational message passing. It is important to note that convergence is not guaranteed when the delta method is used, as the objective function being optimized is no longer a proper bound on the log marginal likelihood. We have experienced divergence in a small number of experiments\footnote{When nonconjugate variational message passing diverges, the lower bound will fluctuate widely and go to negative infinity. In this situation, we can try different initialization values or switch to other methods (e.g. Laplace variational inference or stochastic linear regression).}. Stochastic linear regression \citep{Salimans2013} is useful in such cases as it makes the same assumptions as in nonconjugate variational message passing, but does not require expectations to be evaluated analytically. Instead, updates are obtained stochastically using weighted Monte Carlo by simulating variates from the variational posterior. \cite{Paisley2012} considered an alternate stochastic optimization method for optimizing the intractable lower bound, which uses control variates (functions with tractable expectations that are highly correlated with the intractable function in the lower bound) to reduce the variance of the stochastic search gradient.

To accelerate convergence for large datasets, we develop stochastic variational inference \citep{Hoffman2013} for MMNL models using each of the above three alternatives. In stochastic variational inference, a random subset of agents is selected at each iteration and local variational parameters specific to these agents are optimized. Global variational parameters are then updated using stochastic gradient ascent \citep{Robbins1951}, where the gradients are computed based only on the minibatch of optimized local variational parameters. Stochastic variational inference was developed for conjugate-exponential models and has been applied to latent Dirichlet allocation \citep{Hoffman2010} and the hierarchical Dirichlet process \citep{Wang2011} in topic modeling. \cite{Tan2014} extended stochastic variational inference to logistic and Poisson mixed models using nonconjugate variational message passing. Here, we further extend stochastic variational inference to nonconjugate models via Laplace variational inference and stochastic linear regression, with applications to MMNL models. As large choice sets (e.g. scanner panel data in marketing) become more readily available, stochastic variational inference can play an important role in deriving inference efficiently from large-scale discrete choice models. 

Another contribution of this article is the proposal of a novel strategy to increase minibatch sizes adaptively within stochastic variational inference. At the beginning of the procedure, estimates of the global variational parameters are far from the optimum and only a small minibatch is required to compute the appropriate direction to move in. As the estimates move closer towards the optimum, a more accurate definition of the direction in which to move is required and this can be supplied through using larger minibatches. The idea of adapting batch sizes has been studied by \cite{Orr1996}, \cite{Boyles2011} and \cite{Korattikara2011} in machine learning problems. The results of \cite{Orr1996} are of theoretical interest and they suggest that the best adaptive batch schedule is exponential. \cite{Boyles2011} and \cite{Korattikara2011} construct frequentist hypothesis tests using the Central Limit Theorem for large sums of random variables, and propose increasing batch sizes if the probability of updating parameters in the wrong direction is large. We develop a new criterion based on ``ratio of progress and path" \citep{Gaivoronski1988} while using constant step sizes within the stochastic approximation. Minibatch sizes are increased when the ratio falls beneath a critical value.

This paper is organized as follows. Section \ref{sec_MMNL_model} defines the MMNL model. Section \ref{sec_VI} develops variational inference for the MMNL model using three different approaches. Section \ref{sec_SVI} develops stochastic variational inference for the MMNL model and describes the proposal for increasing minibatch sizes adaptively. Section \ref{sec_assessment} outlines measures for assessing the accuracy of proposed variational methods. Section \ref{sec_examples} considers examples including real and simulated datasets and Section \ref{Conclusion} concludes.

\section{Mixed multinomial logit models of discrete choice} \label{sec_MMNL_model}
The mixed multinomial logit (MMNL) model considered in this paper is defined as follows. Suppose $T_h$ choice events are observed for each agent $h$, $h=1,\dots,H$, and the agent selects from among $J$ alternatives at each choice event. Let the utility that agent $h$ obtains from alternative $j$ at the $t$th choice event be 
\begin{equation*}
U_{htj} =  x_{htj}^T \beta_h + \epsilon_{htj}.
\end{equation*}
Here, $x_{htj}$ is a $K \times 1$ vector of observed variables that relate to alternative $j$ and agent $h$ at the $t$th choice event, $\beta_h$ is a $K \times 1$ random vector of coefficients for agent $h$ representing the agent's preferences and $\epsilon_{htj}$ is a random error term representing unobserved utility. Coefficients in $\beta_h$ are assumed to be distributed as
\begin{equation*}
\beta_h \sim N(\zeta,\Omega) \;\; \text{for} \;\; h=1,\dots, H.
\end{equation*}
Let $y_{ht}=[y_{ht}^1,\dots,y_{ht}^J]^T$ be a $J \times 1$ indicator vector denoting the outcome of agent $h$ at the $t$th choice event and $x_{ht} = [x_{ht1}, \dots,  x_{htJ}]^T$ be a  $J \times K$ matrix of covariates. Assuming that the random error terms $\epsilon_{htj}$ are iid extreme value \cite[see][]{Train2009}, the choice probabilities become
\begin{equation*}
P(y_{ht}^j=1| x_{ht}, \beta_h)= \frac{\exp(x_{htj}^T \beta_h)}{\sum_{j'=1}^J \exp(x_{htj'}^T \beta_h)}
\end{equation*}
for $j=1,\dots,J$, and
\begin{equation*}
p(y_{ht}| x_{ht}, \beta_h) = \prod_{j=1}^J \left\{ \frac{\exp( x_{htj}^T \beta_h)}{\sum_{j'=1}^J \exp( x_{htj'}^T \beta_h)} \right\}^{y_{ht}^j}.
\end{equation*}

We adopt a full Bayesian approach to inference and assume the priors:
\begin{gather}
\zeta |\mu_0,\Sigma_0 \sim N(\mu_0,\Sigma_0), \nonumber \\
\Omega| \nu,a \sim IW(\nu+K-1,\; 2\nu \; \text{diag} (1/a)),  \nonumber \\
\text{where} \; a=[a_1,\dots,a_K]^T,  \label{IWprior}\\
a_k| A_k \stackrel{\text{iid}}{\sim} IG(1/2,\; 1/A_k^2), \; A_k >0 \; \text{for} \; k=1,\dots,K. \label{IGprior}
\end{gather}
The hyperparameters $\mu_0$, $\Sigma_0$, $\nu$ and $A_1,\dots,A_K$ are considered known. The prior distributions for $\Omega$ are marginally noninformative. \cite{Huang2013} showed that \eqref{IWprior} and \eqref{IGprior} induce Half-$t$ distributions on the standard deviation terms in $\Omega$, and a large $A_k$ leads to weakly informative priors on these terms. Moreover, setting $\nu=2$ leads to marginal uniform distributions for all correlation terms in $\Omega$. 

The set of unknown parameters in the MMNL model is $\theta=\{\beta, \zeta, \Omega,\,a\}$, where $\beta = [\beta_1^T, \dots, \beta_H^T]^T$. The variables $\zeta$, $\Omega$ and $a$ are considered as ``global" variables as they are common across all agents. The coefficients in $\beta_h$ are, however, specific to a particular agent $h$ and are considered as ``local" variables. The joint density is given by
\begin{multline*}
p(y,\theta) = \left\{ \prod_{k=1}^K p(a_k|A_k) \right\} p(\Omega| \nu,a) p(\zeta |\mu_0,\Sigma_0)  \\
\times \prod_{h=1}^H  p(\beta_h| \zeta, \Omega)  \prod_{t=1}^{T_h} p(y_{ht}| x_{ht}, \beta_h).
\end{multline*}

\section{Variational inference for the mixed multinomial logit model} \label{sec_VI}
In this section, we develop variational inference for the MMNL model. Three different approaches are presented. The first approach is Laplace variational inference \citep{Wang2013}. The second approach approximates the variational objective function using the multivariate delta method for moments \citep{Bickel2007} and optimization is performed using nonconjugate variational message passing \citep{Knowles2011}. The last approach considers stochastic linear regression \citep{Salimans2013}. We first give a brief introduction to variational methods.

In variational approximation, the true posterior $p(\theta|y)$ is approximated by a more tractable density function $q(\theta)$, which is optimized to be close to $p(\theta|y)$ in terms of the Kullback-Leibler divergence. Minimizing the Kullback-Leibler divergence is equivalent to 
maximizing a lower bound $\mathcal{L}$ on the log marginal likelihood since 
\begin{multline*}
\log p(y) = \int q(\theta) \log \frac{p(y,\theta)}{q(\theta)} \;d\theta+  \int q(\theta) \log \frac{q(\theta)}{p(\theta|y)} \;d\theta, \\
\geq \int q(\theta) \log p(y,\theta) \;d\theta - \int q(\theta) \log q(\theta) \;d\theta = \mathcal{L}.
\end{multline*}

Variational Bayes \citep{Attias1999} assumes $q(\theta)=\prod_{i=1}^m q_i(\theta_i)$ for some partition $\theta= \{\theta_1, \dots, \theta_m\}$, and the optimal $q_i$ maximizing the lower bound $\mathcal{L}$ satisfies
\begin{equation}\label{optimal densities}
q_i(\theta_i) \propto \exp E_{-\theta_i}\{ \log p(y,\theta) \} \;\text{for} \; i=1,\dots,m,
\end{equation}
where $E_{-\theta_i}$ denotes expectation with respect to $\prod_{j \neq i} q_j(\theta_j)$ \citep[see, e.g.][]{Ormerod2010}. When conjugate priors are used, the optimal densities $q_i$ belong to recognizable density families and it suffices to optimize the parameters of each $q_i$.

Applying variational Bayes to the MMNL model, we assume
\begin{equation*}
q(\theta) =q(\zeta)q(\Omega)q(a) \prod_{h=1}^H q(\beta_h).
\end{equation*}
The factors $q(\zeta)$, $q(\Omega)$ and $q(a)$ have conjugate priors. Using \eqref{optimal densities}, the optimal densities can be shown to be $q(\zeta)=N(\mu_\zeta, \Sigma_\zeta)$, $q(\Omega) = IW (\omega, \Upsilon)$ and $q(a) = \prod_{k=1}^K q(a_k)$ where $q(a_k) = IG(b_k,c_k)$ (see \ref{opt_den_deriv}). Let $b=[b_1,\dots,b_K]$ and $c=[c_1,\dots,c_K]$. Variational parameter updates for these factors are given in Algorithm 1. The optimal $q(\beta_h)$ does not belong to any recognizable density family, however, as the likelihood $p(y_{ht}|x_{ht},\beta_h)$ is nonconjugate with respect to the prior over $\beta_h$. Next, we present three approaches on optimizating $q(\beta_h)$.

\subsection{Laplace variational inference}
Suppose $p(\theta|y)$ is some intractable posterior density. Laplace approximation is based on a second order Taylor approximation to $\log p(\theta|y)$, centered at the maximum a posterior (MAP) estimate $\hat{\theta}$ such that
\begin{equation}\label{Laplace}
\log p(\theta|y) \approx \log p(\hat{\theta}|y) + \frac{1}{2}  (\theta-\hat{\theta})^T H(\hat{\theta}) (\theta-\hat{\theta}),
\end{equation}
where $H(\hat{\theta}) = \nabla^2 \log p(\hat{\theta}|y)$. Note that $\nabla \log p(\hat{\theta}|y) =0$ since $ \log p(\theta|y) $ is maximized at $\hat{\theta}$. This gives rise to a Gaussian approximation of the posterior density,
\begin{equation*}
p(\theta|y) \approx N(\hat{\theta}, - H(\hat{\theta})^{-1}).
\end{equation*}

\cite{Wang2013} develop Laplace variational inference for a different class of nonconjugate models by applying Laplace approximation to \eqref{optimal densities}, the optimal density update in variational Bayes. For the MMNL model, \eqref{optimal densities} implies that the optimal $q(\beta_h)$ should satisfy 
\begin{multline*}
q(\beta_h) \propto \exp E_{-\beta_h} \Big\{ \sum_{t=1}^{T_h} \log p(y_{ht}| x_{ht}, \beta_h) \\ 
 + \log p(\beta_h| \zeta,\Omega) \Big\}
\propto \exp \{ f(\beta_h) \}, 
\end{multline*}   
where
\begin{multline} \label{fbetah}
f(\beta_h) = \sum_{t=1}^{T_h} \left[ y_{ht}^Tx_{ht}\beta_h - \log \left\{ \sum_{j=1}^J \exp \left(x_{htj}^T\beta_h \right) \right\} \right]  \\
 -\frac{\omega}{2}(\beta_h-\mu_\zeta)^T  \Upsilon^{-1} (\beta_h -\mu_\zeta).
\end{multline}              
Suppose $f(\beta_h)$ is maximized at $\hat{\beta}_h$.
We consider a second order Taylor approximation of $f(\beta_h)$ at $\hat{\beta}_h$, in a similar fashion as \eqref{Laplace}, such that
\begin{equation} \label{Laplace_approx}
f(\beta_h) \approx f(\hat{\beta}_h) + \frac{1}{2}  (\beta_h-\hat{\beta}_h)^T H(\hat{\beta}_h) (\beta_h-\hat{\beta}_h),
\end{equation}
where $H(\hat{\beta}_h) = \nabla^2 f(\hat{\beta}_h)$. This approximation, combined with $q(\beta_h)\propto \exp \{ f(\beta_h) \}$, results in
\begin{equation*}
q(\beta_h) \approx N(\hat{\beta}_h, - H(\hat{\beta}_h)^{-1}).
\end{equation*}
Let $p_{htj}=\frac{\exp (x_{htj}^T \beta_h)}{\sum_{j'=1}^J \exp(x_{htj'}^T \beta_h)}$ and $p_{ht}=[p_{ht1}, \dots, p_{htJ}]$. The gradient and Hessian of $f$ are given by
\begin{gather}
\nabla f(\beta_h) = \sum_{t=1}^{T_h} x_{ht}^T (y_{ht} - p_{ht}) -\omega \Upsilon^{-1} (\beta_h - \mu_\zeta)  \;\; \text{and} \nonumber  \\
H(\beta_h) =- \sum_{t=1}^{T_h} x_{ht}^T \{  \text{diag}(p_{ht}) - p_{ht}p_{ht}^T \} x_{ht} -\omega \Upsilon^{-1}.  \label{Laplace_grad_Hess}
\end{gather}   
General numerical optimization methods can be used to find $\hat{\beta}_h$. We use the BFGS algorithm via the {\ttfamily optim} function in {\ttfamily R} for optimizing $\hat{\beta}_h$. An approximation $\mathcal{L^*}$ of the variational lower bound $\mathcal{L}$ can be obtained using \eqref{Laplace_approx} and is given in \ref{appendix_Laplace}.

\subsection{Nonconjugate variational message passing with delta method}

The second approach optimizes $q(\beta_h)$ using nonconjugate variational message passing \citep{Knowles2011}. Besides assuming $q(\theta)=\prod_{i=1}^m q_i(\theta_i)$ for some partition $\theta= \{\theta_1, \dots, \theta_m\}$, each $q_i(\theta_i)$ is further assumed to belong to some exponential family such that
\begin{equation*}
q_i(\theta_i)=\exp\{\lambda_i^T t_i(\theta_i)-h_i(\lambda_i)\},
\end{equation*}
where $\lambda_i$ is the vector of natural parameters and $t_i(\cdot)$ are the sufficient statistics. The condition that $\nabla_{\lambda_i}\mathcal{L}=0$ when $\mathcal{L}$ is maximized leads to the fixed point update:
\begin{equation} \label{NCVMP_update}
\lambda_i \leftarrow \text{Cov}_{q_i} [t_i(\theta_i)] ^{-1}\; \nabla_{\lambda_i} E_q\{\log p(y,\theta)\}
\end{equation}
for $i=1,\dots, m$, where $\text{Cov}_{q_i} [t_i(\theta_i)] $ denotes the covariance matrix of $t_i(\theta_i)$. Details are given in \ref{appendix_lb_grad}. Nonconjugate variational message passing thus enables updates of variational parameters to be made in the same spirit as when variational Bayes is applied to conjugate models. There is also flexibility in the evaluation of expectations, such as using bounds or quadrature. However, as a fixed point iterations algorithm, nonconjugate variational message passing is not guaranteed to converge and each update does not necessarily lead to an increase in $\mathcal{L}$. 

Applying nonconjugate variational message passing to the MMNL model, we assume $q(\beta_h)= N(\mu_h,\Sigma_h)$. \cite{Wand2013} showed that for a multivariate Gaussian, the update in \eqref{NCVMP_update} can be simplified to 
\begin{equation} \label{NCVMP Gaussian}
\begin{aligned}
\Sigma_h & \leftarrow -\left[2 \text{vec}^{-1}\left( \nabla\negthinspace_{\text{vec}(\Sigma_h)} E_q\{\log p(y,\theta)\} \right)\right]^{-1} \;\; \text{and} \\ 
\mu_h & \leftarrow \mu_h + \Sigma_h \;  \nabla\negthinspace_{\mu_h} E_q\{\log p(y,\theta)\}.
\end{aligned}
\end{equation}
For a $K \times K$ matrix A, $a = \text{vec}(A)$ is the $K^2 \times 1$ vector obtained by stacking the columns of A under each other, from left to right in order. We let $\text{vec}^{-1}$ denote the reverse operation of $\text{vec}$ so that $A$ can be recovered from $a$ as $A = \text{vec}^{-1}(a)$. The availability of such explicit updates reduces computational cost significantly as $\Sigma_h$ is a full $K \times K$ covariance matrix and numerical optimization of $\Sigma_h$ can be expensive for large $K$. 

For the MMNL model, $E_q\{\log p(y,\theta)\}$ cannot be computed in closed form as
\begin{equation}\label{log_sum_exp}
E_q  \left[\log \left\{ \sum_{j=1}^J \exp(x_{htj}^T \beta_h)    \right\} \right]
\end{equation}
is intractable. Integration using quadrature is computationally intensive, and \cite{Braun2010} approximate \eqref{log_sum_exp} using either Jensen's inequality or the delta method for moments \citep{Bickel2007}. They found that the delta method yielded better performance. Here, we approximate \eqref{log_sum_exp} using the delta method. While \cite{Braun2010} restricted $\Sigma_h$ to be diagonal, a full covariance matrix for $\Sigma_h$ is considered here. This is feasible as optimization using nonconjugate variational message passing is fast. Let $g_t(\beta_h)= \log \left\{ \sum_{j=1}^J \exp(x_{htj}^T \beta_h)  \right\}$. Approximating $g_t(\beta_h)$ with a second order Taylor expansion at $\mu_h$ and taking expectations:
\begin{equation*}
\begin{aligned}
E_q \{ g_t(\beta_h)  \}  &\approx E_q \{ g_t(\mu_h) + \nabla g_t(\mu_h)  (\beta_h - \mu_h) \\
& \quad  + \frac{1}{2} (\beta_h - \mu_h)^T \nabla^2 g_t(\mu_h) (\beta_h - \mu_h)  \}  \\
& = g_t(\mu_h) + \frac{1}{2} \text{tr} \{ \nabla^2 g_t(\mu_h) \Sigma_h \}.
\end{aligned}
\end{equation*}
Let $\rho_{htj}=\frac{\exp (x_{htj}^T \mu_h)}{\sum_{j'=1}^J \exp (x_{htj'}^T \mu_h)}$ and $\rho_{ht} = [\rho_{ht1},\dots, \rho_{htJ}]$. It can be shown that
\begin{multline}\label{delta_approx}
E_q \{ g_t(\beta_h) \} \approx \log \left\{ \sum_{j=1}^J \exp (x_{htj}^T \mu_h) \right\}  \\
+ \frac{1}{2} \text{tr} \left\{ x_{ht}^T \left( \text{diag}(\rho_{ht}) - \rho_{ht} \rho_{ht}^T \right) x_{ht} \Sigma_h  \right\} .
\end{multline}
With \eqref{delta_approx}, updates for $\mu_h$ and $\Sigma_h$ in \eqref{NCVMP Gaussian} can be evaluated in closed form and these are given in Algorithm 1. An approximation $\mathcal{L^*}$ of the variational lower bound $\mathcal{L}$ can also be obtained using \eqref{delta_approx} and details are given in \ref{appendix_delta}. We observe that the delta method leads to good posterior estimation generally. However, this algorithm is not guaranteed to converge as $\mathcal{L}^*$ is not a proper lower bound to the marginal log likelihood $\log p(y)$. An example of such divergence is given in Section \ref{eg_electricity}. In such cases, the third approach described next will be helpful.

\subsection{Stochastic linear regression}

\cite{Salimans2013} present a stochastic linear regression algorithm that allows fixed-form VB to be applied to any posterior (available in closed form up to the proportionality constant) without having to evaluate integrals analytically. Suppose we make the same assumptions as in nonconjugate variational message passing. The fixed point update in \eqref{NCVMP_update} can be expressed as (see \ref{appendix_lb_grad}) 
\begin{equation} \label{SLR} 
\lambda_i = \text{Cov}_{q_i} [t_i(\theta_i)] ^{-1} \;\text{Cov}_{q_i}  \left[ t_i(\theta_i),  E_{-q_i} \{ \log p(y,\theta)\}  \right].
\end{equation}
Instead of evaluating $\text{Cov}_{q_i}  \left[ t_i(\theta_i),  E_{-q_i} \{ \log p(y,\theta)\}  \right]$ and $\text{Cov}_{q_i} [t_i(\theta_i)] $ directly, \cite{Salimans2013} approximate these terms iteratively using weighted Monte Carlo by generating random samples from $q_i(\theta_i)$. When $q_i(\theta_i) = N(\mu_i,\Sigma_i)$, they showed that \eqref{SLR} implies that 
\begin{equation} \label{SLRt}
\Sigma_i= P_i^{-1}  \;\; \text{and} \;\; \mu_i = m_i  + \Sigma_i g_i,
\end{equation}
where 
\begin{gather*}
P_i = - E_{q_i} [ \nabla_{\theta_i}^2 E_{-q_i} \{ \log p(y,\theta)\} ], \\
m_i = E_{q_i}  \{ \theta_i  \}, \\
g_i = E_{q_i} [ \nabla_{\theta_i} E_{-q_i} \{ \log p(y,\theta)\} ],
\end{gather*}
and $\nabla_{\theta_i}^2 E_{-q_i} \{ \log p(y,\theta)\}$ denotes the Hessian matrix of $E_{-q_i} \{ \log p(y,\theta)\}$ in $\theta_i$. \cite{Salimans2013} present a proof for the univariate case by considering a linear transformation and the identities derived by \cite{Minka2001} and \cite{Opper2009}. These identities can be restated as
\begin{multline}\label{Opper}
\nabla\negthinspace_{\mu_i} E_{q_i} \{ V(\theta_i)\} = E_{q_i}\{ \nabla_{\theta_i} V(\theta_i) \} \;\; \text{and} \;\;  \\
\nabla\negthinspace_{\text{vec}(\Sigma_i)} E_{q_i} \{ V(\theta_i)\}= \frac{1}{2}\text{vec} [E_{q_i}\{\nabla_{\theta_i}^2 V(\theta_i) \}],
\end{multline}
where $V(\theta_i)$ is any function in $\theta_i$. We provide a proof for the multivariate case by substituting \eqref{Opper} into \eqref{NCVMP Gaussian}. Details are given in \ref{SLRmG}. The quantities $P_i$, $g_i$ and $m_i$ are approximated stochastically using weighted Monte Carlo. The procedure is described in Figure \ref{SLRproc}.
\begin{figure}[htb]
\centering
\parbox{8cm}{
\hrule \vspace{1mm}
Start with some initialization of $\mu_i$, $\Sigma_i$, $g_i$, $P_i$ and $m_i$. \\
At each iteration $n=1,\dots, N$, 
\begin{itemize}
\item Generate a draw $\hat{\theta}_i$ from $N(\mu_i, \Sigma_i)$.
\item Compute the gradient $\hat{g}_i$ and Hessian $\hat{H}_i$ of $E_{-q_i} \{ \log p(y,\theta)\}$ at $\hat{\theta_i}$.
\item For $0 \leq w \leq 1$, perform the updates 
\begin{gather*}
P_i \leftarrow (1-w) P_i - w \hat{H}_i,\;  g_i \leftarrow(1-w) g_i + w \hat{g}_i \; \; \text{and} \\
m_i \leftarrow (1-w) m_i + w \hat{\theta}_i.
\end{gather*}
\item Compute updates: $\Sigma_i \leftarrow P_i^{-1}$ and $\mu_i \leftarrow m_i + \Sigma_i  g_i$.
\end{itemize}
\hrule}
\caption{\label{SLRproc} Stochastic Linear Regression}
\end{figure}

In Figure \ref{SLRproc}, $q_i = N(\mu_i, \Sigma_i)$ is updated continually and the weights $w$ help to diminish effects from earlier iterations when $q_i$ was less accurate. Following \cite{Salimans2013}, we adopt fixed weights $w$ and average iterates over the second half of the iterations to reduce variability (see Algorithm 2). In setting $N$ (the total number of iterations), it is important to balance accuracy and efficiency. The accuracy of stochastic linear regression deteriorates if $N$ is too small and $\{\mu_i, \Sigma_i\}$ are not sufficiently close to convergence. However, setting $N$ to a very large value can be inefficient as well. When generating draws from $N(\mu_i, \Sigma_i)$, it is computationally more efficient to consider transformation of standard normal random variables using a Cholesky decomposition of $P_i$ instead of evaluating $\mu_i$ and $\Sigma_i$ explicitly at each iteration.

For the MMNL model, the expectation of the log-sum-exp term in \eqref{log_sum_exp} cannot be evaluated in closed form and the delta method was used to approximate this term in the previous section. Stochastic linear regression, on the other hand, does not require expectations to be evaluated analytically and is well suited to the MMNL model. It can also overcome convergence issues in nonconjugate variational message passing, as choosing $w$ to be sufficiently small ensures convergence. 

Instead of updating all variational parameters using stochastic linear regression, a combined approach is considered in this paper. We update $q(\beta_h)$ for $h=1, \dots, H$ using stochastic linear regression while $q(\zeta)$, $q(\Omega)$ and $q(a)$ are updated using explicit variational parameter updates. This approach allows for a straightforward extension to stochastic variational inference, which is discussed in Section \ref{sec_SVI}.

Note that $E_{-q(\beta_h)} \{ \log p(\theta,y) \}=f(\beta_h)$ defined in \eqref{Laplace_approx}. The gradient $g_h$ and Hessian $H_h$ of $E_{-q(\beta_h)} \{ \log p(\theta,y) \}$ are given by $\nabla f(\beta_h)$ and $ H(\beta_h) $ in \eqref{Laplace_grad_Hess}. This result highlights a close connection between stochastic linear regression and Laplace variational inference. While both approximates $q(\beta_h)$ by a Gaussian distribution, an important distinction is that stochastic linear regression optimizes both $\mu_h$ and $\Sigma_h$, while Laplace variational inference optimizes only  $\mu_h$, the location of the Gaussian variational posterior ($\Sigma_h$ is set as the negative inverse Hessian at this point). In the examples, we observe that this procedure in Laplace variational inference often results in underestimation of the standard deviation terms in $\Omega$.

\subsection{Algorithm 1} \label{sec_Alg}
We present the algorithm for computing the variational approximation $q(\theta)$. In Algorithm \ref{Alg1}, $q(\beta_h)$ may be updated using (1) Laplace variational inference, (2) nonconjugate variational message passing or (3) stochastic linear regression. The computational complexity of these algorithms are $\mathcal{O}(HK^2S)$, $\mathcal{O}(HK^2)$ and $\mathcal{O}(HK^2N)$ respectively, where $S$ denotes the number of iterations in BFGS.

\begin{Algorithm*}[htb!]
\centering
\parbox{\textwidth}{
\hrule
\vspace{1mm}
Set $b_k =\frac{\nu+K}{2}$ for $k=1,\dots,K$ and $\omega =H+\nu+K-1$. Initialize $\mu_\zeta=\mu_h=0$, $\Sigma_\zeta = \Sigma_h = 0.01\, I_K$ for $h=1,\dots, H$, $\Upsilon = (\omega -K +1)\, I_K$ and $c=b$. Cycle:
\begin{enumerate}
\item Update $\mu_h$ and $\Sigma_h$ for $h=1,\dots,H$: 
\begin{itemize}
\item \textbf{Laplace variational inference}: Set $\mu_h = \hat{\beta}_h$ and $\Sigma_h=-H(\hat{\beta}_h)^{-1}$, where $\hat{\beta}_h = \argmax_{\beta_h} f(\beta_h)$ and $H(\hat{\beta}_h)$ denotes the Hessian of $f(\beta_h)$ evaluated at $\hat{\beta}_h$. Note: $f(\beta_h)$ and $H(\beta_h)$ are defined in \eqref{Laplace_approx} and \eqref{Laplace_grad_Hess} respectively.
\item \textbf{Nonconjugate variational message passing}: \\
$\Sigma_h \leftarrow \left\{ \sum_{t=1}^{T_h} x_{ht}^T \left( \text{diag}(\rho_{ht}) - \rho_{ht} \rho_{ht}^T \right) x_{ht} + \omega \Upsilon^{-1} \right\}^{-1} $, \\
$\mu_h \leftarrow \mu_h + \Sigma_h \Big[\sum_{t=1}^{T_h}  x_{ht}^T ( y_{ht}-\rho_{ht})
+ x_{ht}^T \left( \text{diag}(\rho_{ht}) - \rho_{ht} \rho_{ht}^T \right) \left\{ x_{ht} \Sigma_h x_{ht}^T \rho_{ht} - \frac{1}{2} \text{diag}(x_{ht} \Sigma_h x_{ht}^T) \right\} - \omega\Upsilon^{-1} (\mu_h-\mu_\zeta) \Big]$.
\item \textbf{Stochastic linear regression}: Initialize $m_h=\mu_h$, $P_h=\Sigma_h^{-1}$, $g_h=0$, $\bar{m}_h=0$, $\bar{P}_h=0$ and $\bar{g}_h=0$. \\
For $n = 1,\dots, N$,
\begin{enumerate}
\item Generate $\hat{\beta}_h$ from $N(\mu_h, \Sigma_h)$.
\item Compute the gradient $\hat{g}_h$ and Hessian $\hat{H}_h$ of $E_{-q(\beta_h)} \{\log p(y,\theta)\}$ at $\hat{\beta}_h$.
\item Set $P_h \leftarrow (1 - w) P_h - w \hat{H}_h$, $g_h \leftarrow(1- w) g_h + w \hat{g}_h$ and $m_h \leftarrow (1 - w) m_h + w \hat{\beta}_h$.
\item Update $\Sigma_h \leftarrow P_h^{-1}$ and $\mu_h \leftarrow \Sigma_h g_h + m_h$.
\item If $n > N/2$, set $\bar{P} _h \leftarrow \bar{P}_h - \frac{2}{N}\hat{H}_h$, $\bar{g} _h \leftarrow \bar{g}_h + \frac{2}{N} \hat{g}_h$ and $\bar{m}_h \leftarrow \bar{m}_h + \frac{2}{N}\hat{\beta}_h$.
\end{enumerate}
Set $\Sigma_h={\bar{P}_h}^{-1}$ and $\mu_h = \Sigma_h \bar{g}_h + \bar{m}_h$.
\end{itemize}
\item $\Sigma_\zeta \leftarrow \left( \Sigma_0^{-1} + H \omega \Upsilon^{-1} \right) ^{-1}$, $\mu_\zeta \leftarrow \Sigma_\zeta \left( \Sigma_0^{-1}\mu_0  +  \omega \Upsilon^{-1} \sum_{h=1}^H \mu_h \right)$.
\item $\Upsilon \leftarrow 2\nu \text{diag}  \left(\frac{b}{c} \right) + \sum_{h=1}^H \{(\mu_h -\mu_\zeta) (\mu_h -\mu_\zeta)^T + \Sigma_h \} + H\Sigma_\zeta$.
\item $c_k \leftarrow \nu \omega \Upsilon_{kk}^{-1} + \frac{1}{A_k^2}$ for $k=1,\dots,K$.
\end{enumerate}
until convergence is reached.
\vspace{1mm}
\hrule}
\caption{Variational inference for MMNL model}\label{Alg1}
\end{Algorithm*} 

In variational algorithms, the lower bound $\mathcal{L}$ is commonly used to check for convergence. However, for stochastic linear regression, it is not easy to compute $\mathcal{L}$ at each iteration. For Laplace approximation and nonconjugate variational message passing, we can only compute approximations of $\mathcal{L}$ which are not guaranteed to increase after each cycle of updates. We consider the following stopping criterion instead. Let $\vartheta = [\mu_\zeta^T, \text{diag}(\Upsilon)^T, c^T]^T$ and $\vartheta_i^{(t)}$ denote the $i$th element of $\vartheta$ at the $t$th iteration. We terminate Algorithm 1 when $\xi^{(t)} = \argmax_i  \frac{| \vartheta_i^{(t+1)} - \vartheta_i^{(t)} |}{|\vartheta_i^{(t)}| } $ is negligible (less than 0.005). For small datasets, there may be some fluctuations in $\xi^{(t)}$ for the option stochastic linear regression. In these cases, we replace $\vartheta$ by its average over the past five iterations.

\section{Stochastic variational inference with adaptive batch sizes} \label{sec_SVI}
In Algorithm 1, the local variational parameters $\mu_h$ and $\Sigma_h$ have to be updated for each agent $h=1, \dots, H$, before the global variational parameters $\mu_\zeta$, $\Sigma_\zeta$, $\Upsilon$ and $c$ can be re-estimated at each iteration. This procedure becomes increasingly inefficient as the number of agents $H$ increases. Stochastic variational inference \citep{Hoffman2013} overcomes this issue by optimizing the global variational parameters using stochastic natural gradient ascent \citep{Robbins1951}. This approach uses only a small random subset of data to compute unbiased estimates of the natural gradients at each iteration, and computation time is reduced significantly when $H$ is large. The procedure is described in Figure \ref{svi}. As large datasets in discrete choice modeling become increasingly common, stochastic variational inference can play an important role in estimating MMNL models. 
\begin{figure}[htb]
\hrule\vspace{1mm}
At each iteration,
\begin{itemize}
\item draw a minibatch $B$ of agents randomly from the entire pool of agents.
\item Optimize local variational parameters $\mu_h$ and $\Sigma_h$ for agents $h \in B$ (as a function of the global variational parameters at their current setting).
\item Update global variational parameters using stochastic natural gradient ascent. Noisy gradient estimates are computed using optimized local variational parameters $\mu_h$ and $\Sigma_h$ for agents $h \in B$.
\end{itemize}   
\hrule\vspace{1mm}
\caption {\label{svi} Stochastic variational inference procedure}
\end{figure}

We develop stochastic variational inference for the MMNL model by building upon the methods discussed in Section \ref{sec_VI}. The use of Laplace variational inference, nonconjugate variational message passing and stochastic linear regression within stochastic variational inference are explored. In addition, a novel approach to increase minibatch sizes adaptively is proposed. First, we explain how global variational parameters are updated.

\subsection{Stochastic gradient ascent updates}
In stochastic variational inference, global variational parameters are updated using stochastic natural gradient ascent. At the $l$th iteration, an update of the form 
\begin{equation}\label{stoc_updates}
\lambda_i^{(l+1)} = \lambda_i^{(l)} + \alpha_l \; \tilde{\nabla}_{\lambda_i} \mathcal{L}
\end{equation}
is applied where $\alpha_l$ represents a small step taken in the direction of $\tilde{\nabla}_{\lambda_i} \mathcal{L}$, the natural gradient of the lower bound with respect to $\lambda_i$. In stochastic natural gradient ascent, noisy estimates are used in place of the true natural gradients. \cite{Hoffman2013} provides a motivation for the use of natural gradients in coordinate ascent by considering the geometry of the parameter space. The natural gradient $\tilde{\nabla}_{\lambda_i} \mathcal{L}$ can be obtained by premultiplying the ordinary gradient $\nabla_{\lambda_i} \mathcal{L}$ with the inverse of the Fisher information matrix of $q_i(\theta_i)$ \cite{Amari1998}. When $q_i(\theta_i)$ belongs to an exponential family, the natural gradient is given by (see \ref{appendix_lb_grad})
\begin{equation} \label{nat_grad}
\tilde{\nabla}_{\lambda_i} \mathcal{L} =  \text{Cov}_{q_i} [t_i(\theta_i)] ^{-1}\; \nabla_{\lambda_i} E_q\{\log p(y,\theta)\} - \lambda_i.
\end{equation}

Let $\lambda_\zeta$, $\lambda_\Omega$ and $\lambda_{\beta_h}$ denote the natural parameter vectors of $q(\zeta)$, $q(\Omega)$ and $q(\beta_h)$ respectively. Let $\lambda_{\beta_h}^{\text{opt}}$ denote $\lambda_{\beta_h}$ optimized as a function of the current global variational parameters. From \eqref{nat_grad},
\begin{multline*}
\tilde{\nabla}_{\lambda_\zeta} \mathcal{L} =  \text{Cov}_{q(\zeta)} [t(\zeta)] ^{-1} \nabla_{\lambda_\zeta} \bigg[ E_q\{\log p(\zeta| \mu_0, \Sigma_0) \} \\ 
+ \sum_{h=1}^H  E_q \{\log p(\beta_h|\zeta, \Omega)\}\vert_{\lambda_{\beta_h} = \lambda_{\beta_h}^{\text{opt}}} \bigg] - \lambda_\zeta.
\end{multline*}
Suppose that a minibatch $B$ of agents is drawn randomly from the entire pool of agents. An unbiased estimate of $\tilde{\nabla}_{\lambda_\zeta} \mathcal{L}$ is 
$\hat{\lambda}_\zeta - \lambda_\zeta$, where
\begin{multline*}
 \hat{\lambda}_\zeta = \text{Cov}_{q(\zeta)} [t(\zeta)] ^{-1} \nabla_{\lambda_\zeta} \bigg[ E_q\{\log p(\zeta| \mu_0, \Sigma_0) \} +  \\ \frac{H}{|B|}\sum_{h \in B}  E_q \{\log p(\beta_h|\zeta, \Omega)\}\vert_{\lambda_{\beta_h} = \lambda_{\beta_h}^{\text{opt}}} \bigg].
\end{multline*}
Similarly, an unbiased estimate of $\tilde{\nabla}_{\lambda_\Omega} \mathcal{L}$ is $\hat{\lambda}_\Omega - \lambda_\Omega$, where
\begin{multline*}
 \hat{\lambda}_\Omega = \text{Cov}_{q(\Omega)} [t(\Omega)] ^{-1} \nabla_{\lambda_\Omega} \bigg[E_q\{\log p(\Omega| \nu, a) \} + \\ \frac{H}{|B|}\sum_{h \in B}  E_q \{\log p(\beta_h|\zeta, \Omega)\}\vert_{\lambda_{\beta_h} = \lambda_{\beta_h}^{\text{opt}}} \bigg].
\end{multline*}
From \eqref{stoc_updates}, the stochastic gradient updates for $\lambda_\zeta$ and $\lambda_\Omega$ thus take the form of 
\begin{equation*}
\begin{aligned}
\lambda_\zeta^{(l+1)} &= (1- \alpha_l ) \, \lambda_\zeta^{(l)} + \alpha_l  \, \hat{\lambda}_\zeta  \;\; \text{and} \\
\lambda_\Omega^{(l+1)} &= (1- \alpha_l ) \, \lambda_\Omega^{(l)} + \alpha_l  \, \hat{\lambda}_\Omega.
\end{aligned}
\end{equation*}
The present estimate $\lambda_\zeta^{(l+1)}$ is a weighted average of the previous estimate $\lambda_\zeta^{(l)}$ and the estimate of $\lambda_\zeta$ computed using minibatch $B$, $\hat{\lambda}_\zeta$. Simplified updates are given in Algorithm 2. Note that updates in Algorithm 1 are recovered when $|B|=H$ and $\alpha_l=1$. The updates for $c$ remain the same as in Algorithm 1 as they do not depend on the local variational parameters. 

The iterates can be shown to converge under certain regularity conditions \citep[see][]{Spall2003}. In particular, the stepsizes $\alpha_l$ should satisfy
\begin{equation*}
\alpha_l\rightarrow 0,\;\;\;\sum_{l=0}^\infty \alpha_l=\infty,\;\;\;\text{and}\;\;\;\sum_{l=0}^\infty \alpha_l^2<\infty.
\end{equation*}
A commonly used gain sequence that satisfies these rules is $\alpha_l = \frac{d}{(l+D)^\gamma}$, where $0.5< \gamma \leq 1$. Smaller values of $\gamma$ slow down the rate at which stepsizes decline, $d > 1$ helps to maintain larger stepsizes in later iterations and $D \geq 0$ is a stability constant that helps avoid unstable behavior in early iterations. The performance of stochastic approximation algorithms tends to be very sensitive to the rate of decrease of stepsizes and some tuning is usually required to achieve optimal performance. A review of rules for choosing stepsizes in deterministic or stochastic manners can be found in \cite{Powell2011}. \cite{Ranganath2013} developed an adaptive stepsize for stochastic variational inference, which is designed to minimize the expected distance between stochastic and batch updates.

\subsection{Using adaptive batch sizes} \label{sec_adaptive_batch_sizes}
We propose a new approach towards constructing an automatic algorithm for implementing stochastic variational inference. In contrast to existing approaches of keeping the minibatch size fixed and using a stepsize with a decreasing trend (deterministic or adaptive) to reduce noise, we propose increasing the minibatch size adaptively as optimization proceeds, until the minibatch size is equal to the size of the whole dataset. In cases where the dataset is too large to be processed in batch mode, our adaptive strategy may still be useful with the upper bound set at a feasible minibatch size. The idea of increasing batch size adaptively has been investigated by \cite{Orr1996}, \cite{Boyles2011} and \cite{Korattikara2011} in machine learning tasks. 

Intuitively, estimates of the global variational parameters are far from the optimum at the beginning and hence only a small minibatch is required to compute the appropriate direction to move in. As the estimates move closer towards the optimum, a more accurate definition of the direction in which to move is required and this can be supplied through using larger minibatches. Eventually, the entire dataset is used\footnote{This approach requires that the gradient for all observations be computable eventually.}. This ensures convergence and the same level of accuracy can also be attained as in batch mode. With this approach, we avoid having to specify a stopping criterion for a stochastic approximation algorithm. Developing a good stopping criterion can be very challenging. Most commonly used criteria do not guarantee that the terminal iterate is close to the optimum and may be satisfied by chance \citep{Jank2006}. Very often, stochastic approximation algorithms are terminated based on some predetermined computational budget \citep{Hoffman2013, Ranganath2013}. The risk of ``apparent convergence'' associated with a declining stepsize is also avoided. ``Apparent convergence'' refers to the case where iterates appear to have converge due to diminishing stepsizes even though they are actually far from the optimum \citep[see][]{Powell2011}. 

To obtain maximal computational savings, the minibatch size should be increased only when the current minibatch size can no longer provide adequate information about the appropriate direction in which to move. \cite{Orr1996} investigates the convergence behavior of least mean squares and derives a formula for the optimal minibatch size at each iteration (by maximizing the reduction in weight error per input presented). \cite{Orr1996} notes that their results are of interest theoretically but difficult to apply in practice due to the presence of complex quantities such as the Hessian, which are hard to compute. \cite{Boyles2011} and \cite{Korattikara2011} construct frequentist hypothesis tests to determine if parameter updates are likely to be in the correct direction, and suggest increasing the minibatch size by a certain factor, when all parameters are failing their hypothesis tests. They observe that stochastic gradients in gradient ascent algorithms often involve averaging over a large number of random variables and make use of the Central Limit Theorem as a basis for their tests. We have attempted to apply their approach in stochastic variational inference. However, we find that in our context, the hypothesis tests tend to fail at a stage which is still too early for minibatch sizes to be increased, resulting in suboptimal performance.

\subsection{Proposed strategy} \label{sec_proposed strategy}
We propose the following strategy for increasing minibatch sizes adaptively. Starting with a minibatch $B$, we implement the procedure in Figure \ref{svi} repeatedly, updating the global variational parameters with a constant stepsize. In general stochastic gradient optimization algorithms, constant stepsizes are popular even though they do not lead to formal convergence as the algorithm tends to be more robust \cite[allowing stepsizes to decrease too quickly can reduce the rate of convergence of the algorithm and produce ``apparent convergence",][]{Powell2011}. With constant stepsizes, iterates tend to move monotonically towards the optimum at first. However, near the optimum, they will bounce around instead of converge towards it as stepsizes remain large. This oscillating phenomenon is an indication that the current minibatch size is no longer adequate in defining the direction to move. More resolution is required and we increase the minibatch size by a factor $\kappa$. This process is repeated until the whole dataset is used. 

To detect if iterates have reached the stage where they are merely bouncing around the optimum, we consider the ``ratio of progress and path" defined in \cite{Gaivoronski1988} as
\begin{equation*}
\phi^{(l)} = \frac{| \lambda_i^{(l-M)} - \lambda_i^{(l)}|}{\sum_{r=l-M}^{l-1} | \lambda_i^{(r)} - \lambda_i^{(r+1)}| }
\end{equation*}
for a univariate variable $\lambda_i$ at iteration $l$. \cite{Gaivoronski1988} used this ratio to define an adaptive stepsize which decreases by a factor if $\phi^{(l)}$ is less than a certain value and remains the same otherwise. The ratio $\phi^{(l)}$ lies between zero and one. It is zero when $ \lambda_i^{(l-M)} = \lambda_i^{(l)}$ (i.e. there is no progress after $M$ iterations) and it is one when the path from $\lambda_i^{(l-M)}$ to $\lambda_i^{(l)}$ is monotonic. Small values of $\phi^{(l)}$ indicate that the path of the algorithm is erratic and there is a lot of back and forth movement. This ratio is thus a good indicator of whether iterates are close to the optimum and are simply bouncing around. Note that $\lambda_i^{(l-M)}, \dots, \lambda_i^{(l)}$ will have to be stored in memory for the computation of $\phi^{(l)}$. 

We monitor the ``ratio of progress and path" for elements in $\mu_\zeta$ and the diagonal of $\Upsilon$. In the examples, we set $M=20$ and store the past $M$ values of $\mu_\zeta$ and $\text{diag}(\Upsilon)$. However, as it is unlikely that the algorithm will have to stay at every minibatch size for more than 20 iterations, we start computing the ratios as soon as $l >5$ using the available history. Thus, we compute 
\begin{equation*}
\begin{aligned}
\phi^{(l)}_{1k} &=  \left\{
  \begin{array}{l l}
 \frac{| \Upsilon_{kk}^{(0)} - \Upsilon_{kk}^{(l)}|}{\sum_{r=0}^{l-1} |\Upsilon_{kk}^{(r)} - \Upsilon_{kk}^{(r+1)}|} \;\;\; \text{if $5  < l < M$}\\ [4mm]
 \frac{| \Upsilon_{kk}^{(l-M)} - \Upsilon_{kk}^{(l)}|}{\sum_{r=l-M}^{l-1} | \Upsilon_{kk}^{(r)} - \Upsilon_{kk}^{(r+1)}| } \;\;\; \text{if $l \geq M$}
  \end{array} \right.
\text{and}\;\; \\
 \phi^{(l)}_{2k} &=  \left\{
   \begin{array}{l l}
 \frac{| {\mu_\zeta}_k^{(0)} - {\mu_\zeta}_k^{(l)}|}{\sum_{r=0}^{l-1} |{\mu_\zeta}_k^{(r)} - {\mu_\zeta}_k^{(r+1)}| }  \;\;\; \text{if $5  < l < M$}\\ [4mm]
  \frac{|{\mu_\zeta}_k^{(l-M)} -{\mu_\zeta}_k^{(l)}|}{\sum_{r=l-M}^{l-1} |{\mu_\zeta}_k^{(r)} - {\mu_\zeta}_k^{(r+1)}|  } \;\;\; \text{if $l \geq M$}.
   \end{array} \right.
\end{aligned}
\end{equation*}
The minibatch size is increased by a factor $\kappa$ when the minimum value of the $2K$ ratios falls beneath a critical value $\Phi$. We allow $\Phi$ to vary according to the minibatch size $|B|$. For a small $|B|$, a smaller $\Phi$ is required as the path of the algorithm can be quite erratic even though progress is being made due to the greater randomness present between iterations.

\subsection{Algorithm 2}

The proposed algorithm for implementing stochastic variational inference using adaptive batch sizes is outlined in Algorithm 2. 

\begin{Algorithm*}
\centering
\parbox{0.8\textwidth}{
\hrule
\vspace{1mm}
Set $b_k =\frac{\nu+K}{2}$ for $k=1,\dots,K$ and $\omega =H+\nu+K-1$. Initialize $\mu_\zeta=\mu_h=0$, $\Sigma_\zeta = \Sigma_h = 0.01\, I_K$ for $h=1,\dots, H$, $\Upsilon = (\omega -K +1)\, I_K$, $c=b$, $l=0$ and $|B|=25$. \\
While $|B| < H$,
\begin{enumerate}
\item $l \leftarrow l+1$
\item Randomly select a minibatch $B$ of $|B|$ agents from the entire pool of agents.
\item Optimize $\mu_h$ and $\Sigma_h$ for $h \in B$ using  
\begin{itemize}
\item Laplace variational inference (as in Algorithm 1),
\item nonconjugate variational message passing (Perform updates in Algorithm 1 repeatedly until convergence is reached), or
\item stochastic linear regression (as in Algorithm 1)
\end{itemize}
\item $\Sigma_\zeta \leftarrow \left( \Sigma_0^{-1} + H \omega \Upsilon^{-1} \right) ^{-1}$, $\mu_\zeta \leftarrow (1-\alpha_{|B|})\mu_\zeta + \alpha_{|B|} \Sigma_\zeta \left( \Sigma_0^{-1}\mu_0  +  \omega \Upsilon^{-1} \frac{H}{|B|}\sum_{h \in B} \mu_h \right)$.
\item $\Upsilon \leftarrow (1 - \alpha_{|B|}) \Upsilon +\alpha_{|B|}  \left[ 2\nu \text{diag}  \left(\frac{b}{c} \right) + \frac{H}{|B|}\sum_{h \in B} \{(\mu_h -\mu_\zeta) (\mu_h -\mu_\zeta)^T + \Sigma_h \} + H\Sigma_\zeta \right]$.
\item $c_k \leftarrow \nu \omega \Upsilon_{kk}^{-1} + \frac{1}{A_k^2}$ for $k=1,\dots,K$.
\item If $l >5 $, compute $\phi_{1k}^{(l)}$ and $\phi_{2k}^{(l)}$ for $k=1,\dots, K$. \\ [1mm]
If min $\left\{\phi_{1k}^{(l)},\, \phi_{2k}^{(l)} \,|\, k=1,\dots,K \right\} < \Phi_{|B|}$, $|B| \leftarrow \text{min}\; \{\kappa|B|, H\}$, $l \leftarrow 0$.
\end{enumerate}
If $|B|=H$, cycle
\begin{enumerate}
\item Update $\mu_h$ and $\Sigma_h$ for $h =1,\dots,H$ using  
\begin{itemize}
\item Laplace approximation (as in Algorithm 1),
\item nonconjugate variational message passing (Perform updates in Algorithm 1 repeatedly until convergence is reached in the first iteration and just once subsequently), or
\item stochastic linear regression (as in Algorithm 1)
\end{itemize}
\item $\Sigma_\zeta \leftarrow \left( \Sigma_0^{-1} + H \omega \Upsilon^{-1} \right) ^{-1}$, $\mu_\zeta \leftarrow \Sigma_\zeta \left( \Sigma_0^{-1}\mu_0  +  \omega \Upsilon^{-1} \sum_{h=1}^H \mu_h \right)$.
\item $\Upsilon \leftarrow  2\nu \text{diag}  \left(\frac{b}{c} \right) + \sum_{h=1}^H \{(\mu_h -\mu_\zeta) (\mu_h -\mu_\zeta)^T + \Sigma_h \} + H\Sigma_\zeta $.
\item $c_k \leftarrow \nu \omega \Upsilon_{kk}^{-1} + \frac{1}{A_k^2}$ for $k=1,\dots,K$.
\end{enumerate}
until convergence.
\vspace{1mm}
\hrule}
\caption{Stochastic variational inference for MMNL model using adaptive batch sizes}\label{Alg2}
\end{Algorithm*}

In stochastic variational inference, the local variational parameters should be optimized as a function of the global variational parameters at their current setting (step 3 of Algorithm 2). Laplace variational inference optimizes only $\mu_h$, the location of the Gaussian variational posterior and sets $\Sigma_h$ as the inverse of the negative Hessian at this point. Using our adaptive batch size approach, convergence is ensured, however, as the entire dataset is used eventually. For nonconjugate variational message passing, the updates for $\mu_h$ and $\Sigma_h$ are recursive and they have to be performed repeatedly until convergence is reached in order for $\mu_h$ and $\Sigma_h$ to be optimized. Let $\mu_B^{(l)}$ be a concatenation of the vectors $\mu_h$ for $h \in B$ at the $l$th iteration. In the examples, we terminate the number of iterations in nonconjugate variational message passing when $||\mu_B^{(l)} - \mu_B^{(l-1)}||/||\mu_B^{(l)}|| < 0.1$ where $||\cdot||$ denotes the Euclidean norm, or when the number of iterations hit a maximum of three. We have used a loose stopping criterion here for greater computational efficiency. In the case of stochastic linear regression, we fix the number of iterations at $N$ for simplicity and assume that this number of iterations is sufficient for $\{ \mu_h, \Sigma_h\}$ to be sufficiently close to convergence. 

As described in Section \ref{sec_proposed strategy}, constant stepsizes are used within each minibatch size, and we allow the stepsize to increase with the minibatch size $|B|$. Intuitively, smaller stepsizes are required at the beginning as we are less confident in the direction of gradient ascent computed based on the small minibatch of optimized variational parameters. As the minibatch size increase, our confidence level increases. The stepsize is 1 when the algorithm transits to batch mode (when $|B|=H$). In the examples, we start with a minibatch size of $|B|=25$ and an initial stepsize $\alpha_{|B|}=0.4$. We let the stepsize increase linearly with the minibatch size\footnote{We let $\alpha_{|B|} = 0.4 +0.6(|B|-25)/(H-25)$, and $\phi_{|B|}$ is computed similarly.} until it reaches 1 when $|B|=H$. We find that the performance of minibatches with sizes smaller than 25 tend to be more erratic. The critical value $\Phi_{|B|}$ for the ``ratio of progress and path'' is also initialized at 0.4 for the initial minibatch size of 25, and allowed to increased linearly with the minibatch size $|B|$ until it reaches 1 when $|B|=H$. We have used linear increments as this is a straightforward option. It is possible to experiment with other settings. Our experiments indicate that minor variations from these settings do not result in much changes in the performance of Algorithm 2.

\section{Assessment of proposed variational methods} \label{sec_assessment}
The standard Bayesian procedure for obtaining inference from MMNL models is via MCMC methods. \cite{Train2009} describes how posterior samples for a MMNL model can be obtained using a Metropolis-Hastings within Gibbs algorithm. \cite{Rossi2005} proposed an improved random walk Metropolis algorithm for drawing $\beta_h$  using a fractional likelihood approach. They demonstrate that the improved random walk Metropolis exhibits better mixing and dissipates initial conditions in a shorter time than a random walk Metropolis and an independence Metropolis sampler. This algorithm in implemented in the {\ttfamily R} package {\ttfamily bayesm} via the function {\ttfamily rhierMnlRwMixture}. We modify this function slightly to accommodate the marginally noninformative priors for $\Omega$ defined in \eqref{IWprior} and \eqref{IGprior}, and use it to compare MCMC with proposed variational methods.

\subsection{Predictive choice distribution}
For assessing the accuracy of proposed variational methods, we use the measures discussed in \cite{Braun2010}, which are based on the predictive choice distribution. The true predictive choice distribution of a $J \times 1 $ vector of outcomes $y_{\text{new}}$ given the $J \times K$ matrix of observed variables $x_{\text{new}}$ is defined as 
\begin{equation} \label{pcd}
p_{\text{true}}(y_{\text{new}}|x_{\text{new}}, \zeta, \Omega) = \int p(y_{\text{new}}|x_{\text{new}},\beta) p(\beta|\zeta,\Omega) \;d\beta. 
\end{equation}
For data simulated artificially from the MMNL model, the true predictive choice distribution can be computed using Monte Carlo integration as $\zeta$ and $\Omega$ are known. In Section \ref{eg_sim}, we use 1000,000 draws of $\beta$ from $N(\beta| \zeta,\Omega)$ to compute the true predictive choice distribution for simulated data. \cite{Braun2010} showed that variability arising from Monte Carlo integration is not noticeable if this many draws of $\beta$ are used. 

A point estimate of the predictive choice distribution can be obtained by taking the mean of \eqref{pcd} under the posterior of $\zeta$ and $\Omega$:
\begin{multline}\label{pcd_est}
\hat{p}(y_{\text{new}}|x_{\text{new}}, y) = \int \left \{\int p(y_{\text{new}}|x_{\text{new}},\beta) \right. \\
\left. \times p(\beta|\zeta,\Omega) \;d\beta \right\} p(\zeta,\Omega|y) \; d\zeta \, d\Omega.
\end{multline}
The estimated predictive choice distribution can be computed using Monte Carlo integration for both variational and MCMC methods. For variational methods, we approximate the posterior density $p(\zeta,\Omega|y)$ with the fitted variational posterior density $q(\zeta) q(\Omega)$. We use 500 draws of $\{  \zeta,\Omega\}$ from $q(\zeta) q(\Omega)$ for variational methods and 10,000 draws for MCMC. More samples are used in the case of MCMC as there is some autocorrelation among the draws. For the estimated predictive choice distributions, we use 10,000 iid draws of $\beta$.

Following \cite{Braun2010}, we use the total variation (TV) metric to compute the distance between two predictive choice distributions \citep[see, e.g.][]{Levin2009}. For the simulated datasets, the TV distance between the estimated and true predictive choice distributions at the attribute matrix $x_{\text{new}}$ can be computed using
\begin{multline*}
\text{TV}[p_{\text{true}}(y_{\text{new}}|x_{\text{new}}), \hat{p}(y_{\text{new}}|x_{\text{new}}) ] \\
= \frac{1}{2}\sum_{j=1}^J |p_{true} (y_{\text{new}}^j=1| x_{\text{new}}) - \hat{p} (y_{\text{new}}^j=1| x_{\text{new}})|.
\end{multline*}
For real datasets, the true predictive choice distribution is unknown and we compute the TV distances between the predictive choice distribution estimated using MCMC and the variational methods instead. This provides a means of assessing the degree of agreement between MCMC and variational methods.

\section{Examples} \label{sec_examples}
In the following examples, the performances of Laplace approximation (Laplace), nonconjugate variational message passing (NCVMP) and stochastic linear regression (SLR) are compared with that of MCMC in terms of the predictive choice distribution. We set $N=40$ and $w=0.25$ for SLR in both Algorithms 1 and 2. The runtime of SLR in Algorithm 1 varies slightly between runs and we report the mean runtime and standard deviation over 5 runs. Estimates of the variational parameters are almost identical in each run of SLR. We use the run with runtime closest to the mean runtime to compute the predictive choice distribution. For Algorithm 2, there is greater variation and we repeat runs for each alternative ten times. The mean runtime and standard deviation over the ten runs are reported.

For MCMC, 4 independent chains were run in each example and the first half of each chain was discarded as burn-in. In each example, 10,000 draws remained after thinning and the Gelman-Rubin diagnostics were used to check that these draws are a good approximation of the posterior distribution. These draws were then used to compute the estimated predictive choice distribution for MCMC. We note that there are some inherent difficulties in comparing runtimes of variational methods with MCMC. MCMC is a simulation-based method and runtimes depend on the thinning factor, length of burn-in and number of sampling iterations. These are problem dependent; a higher thinning factor or longer burn-in may be required when mixing is poor. On the other hand, variational Bayes is deterministic and the time to convergence (which is also problem dependent) depends on the initialization and stopping rule. Due to these concerns, we only present the results of a predictive log-likelihood experiment in Section 6.1, where the variational algorithms and MCMC are given an equal computational budget.

We use a vague $N(0,10^6)$ for $\zeta$ and set $\nu = 2$, $A_k= 10^3$ for $k=1, \dots, K$. All code was written in {\tt R} and computations were carried out on a 64-bit 3.20 GHz Intel Core i5 processor with 8 GB of memory. Our R code is available as supplementary materials for this article.

\subsection{Vehicles choice stated preference experiments}
The first example considers data representing consumers' choices among vehicles in stated preference experiments. This dataset consists of $H=100$ respondents and comes from a study for Toyota and General Motors on the marketability of electric and hybrid vehicles. It is available at \url{http://elsa.berkeley.edu/users/train/ec244.html} and is a subset of the full dataset of 500 respondents \citep[see][]{Hess2006}.

Each respondent faced up to 15 experiments and chose among $J=3$ different vehicles in each experiment. The vehicles are described in terms of the following attributes: negative of price (in ten thousand dollars), negative of operating cost per month (in ten dollars), engine type (gas, electric, or hybrid), range in hundreds of miles between recharging (if engine is electric) and performance level (high, medium, or low). The range of a vehicle is set to zero for all non-electric vehicles. An indicator variable for hybrid vehicle is included as a covariate. Performance level is represented using two dummy variables with low performance as the base. We have $K=6$ covariates and $10 \leq T_h \leq 15$ as some respondents did not complete all 15 experiments.

We run Algorithm 1 using Laplace, NCVMP and SLR and runtimes are given in Table \ref{vehicle_runtime}. MCMC was run for 4 chains, each with 25000 iterations and a thinning factor of 5 was applied. Mixing is poor for parameters corresponding to the indicator variable for medium performance and so a larger number of iterations and higher thinning factor were used. 
\begin{table}[htb!]
\centering\ra{1.1}
\begin{tabular}{@{}lcccc@{}}
\hline
             & Laplace  & NCVMP & SLR   \\  \hline  
Times  & 19          & 4             & 25 (1)  \\  \hline
\end{tabular}
\caption{\label{vehicle_runtime} Vehicles example: CPU times (seconds) for Algorithm 1 (Laplace, NCVMP and SLR). Standard deviation over repeated runs given in brackets.}
\end{table}

\begin{table*}[htb!]
	\centering\ra{1.1}
	\begin{tabular}{@{}lcccccc@{}}
		\hline
		& Min.     & 1st Qu.      & Median       & Mean       & 3rd Qu.   & Max  \\  \hline  
		Laplace vs. MCMC  & 0.08 \%   & 1.36 \%      & 2.14 \%      & 2.15 \%    & 2.90 \%  & 4.60 \%   \\ 
		NCVMP vs. MCMC & 0.02 \%   & 0.21 \%      & 0.32 \%      & 0.34 \%    & 0.46 \%  & 0.87 \%   \\
		SLR vs. MCMC       & 0.01 \%   & 0.19 \%      & 0.30 \%      & 0.34 \%    & 0.45 \%  & 1.10 \%    \\  \hline
	\end{tabular}
	\caption{\label{vehicle_TV} Vehicles example: Summary of 400 TV distances between predictive choice probabilities computed using MCMC and Algorithm 1 (Laplace, NCVMP and SLR).}
\end{table*}
Table \ref{vehicle_TV} provides a summary of the TV distances between the predictive choice distribution estimated by MCMC and the variational methods. The TV distances were computed at 400 attribute matrices, obtained by randomly selecting four choice event covariate matrices $x_{ht}$ from each of the 100 respondents. The performances of NCVMP and SLR are almost indistinguishable, and both performed significantly better than Laplace approximation. NCVMP and SLR are actually very similar; the main difference is that SLR does not require expectations to be evaluated analytically while NCVMP used the delta method to approximate expectations of the log-sum-exp terms. However, due to the use of simulations and weighted Monte Carlo, SLR is slower than NCVMP. Laplace approximation, on the other hand, considers a second order Taylor approximation within the coordinate ascent update of variational Bayes. This approximation does not seem to work well in the context of MMNL models and the performance of Laplace approximation is worse than NCVMP and SLR in all examples. However, unlike NCVMP, which is not guaranteed to converge, Laplace approximation is very stable and no convergence issues were encountered in any example. The TV distances of NCVMP and SLR from MCMC are very small with an average of 0.3\% and a maximum of around 1\%, showing very good agreement.

We also perform a five-fold cross-validation experiment to assess the predictive performance of variational methods and MCMC given equal computational budget. We randomly divide the 100 respondents in the dataset into five folds; each fold is used in turn as a test set and the remaining folds are used for training. For each fold, we use the model estimated from the training data to compute the predictive log-likelihood of the held out test set $\mathcal{T}_i$ as 
\begin{equation*}
\sum_{h \in \mathcal{T}_i} \sum_{t=1}^{T_h} \sum_{j=1}^J y_{ht}^j \log \hat{p}(y_{ht}^j|x_{htj},y_{\text{train}}),
\end{equation*}
where $\hat{p}(y_{ht}^j|x_{htj},y_{\text{train}})$ is estimated as in Section 5.1. In this experiment, the variational methods were run to convergence and MCMC was allocated the same amount of time that SLR (slowest) took to converge. Within this duration, MCMC was able to complete approximately 1200 iterations. We discard the first half of the iterations and use only the second half to compute the predictive log-likelihood. Predictive log-likelihoods and runtimes averaged over five folds are given in Table \ref{vehicle_pred}. Under the same time constrains, predictive performance of NCVMP and SLR are better than MCMC but Laplace is doing worse than MCMC. It is worth noting that NCVMP used only $\frac{1}{7}$ of the budget and yet attained the best predictive performance.
\begin{table}[htb!]
\centering\ra{1.1}
\begin{tabular}{@{}lcc@{}}
\hline & Predictive log-likelihood & Times \\ \hline
Laplace & -282.44  & 15 \\
NCVMP &  -281.58 & 3 \\
SLR &  -281.71 & 21 \\
MCMC  & -281.85 & 21 \\ \hline
\end{tabular}
\caption{\label{vehicle_pred} Vehicles example: Predictive log-likelihoods and runtimes averaged over five folds for Algorithm 1 (Laplace, NCVMP and SLR) and MCMC.}
\end{table}

\subsection{Simulated data} \label{eg_sim}
In this simulation study, we generate two datasets from the MMNL model in Section \ref{sec_MMNL_model}. One is of low heterogeneity with $\Omega=0.25 I_K$ while the other is of high heterogeneity with $\Omega=I_K$. In each dataset, there are $H=10,000$ agents, $J=12$ alternatives, $K=10$ attributes and $T_h=25$ observed events for each agent $h$. $\zeta$ consists of equally spaced values from $-2$ to 2. Entries in the attribute matrices $x_{ht}$ were generated independently from $N(0, 0.5^2)$. This set-up is similar to that in \cite{Braun2010}.

\begin{figure}[htb!]
\centering
\includegraphics[width=0.238\textwidth]{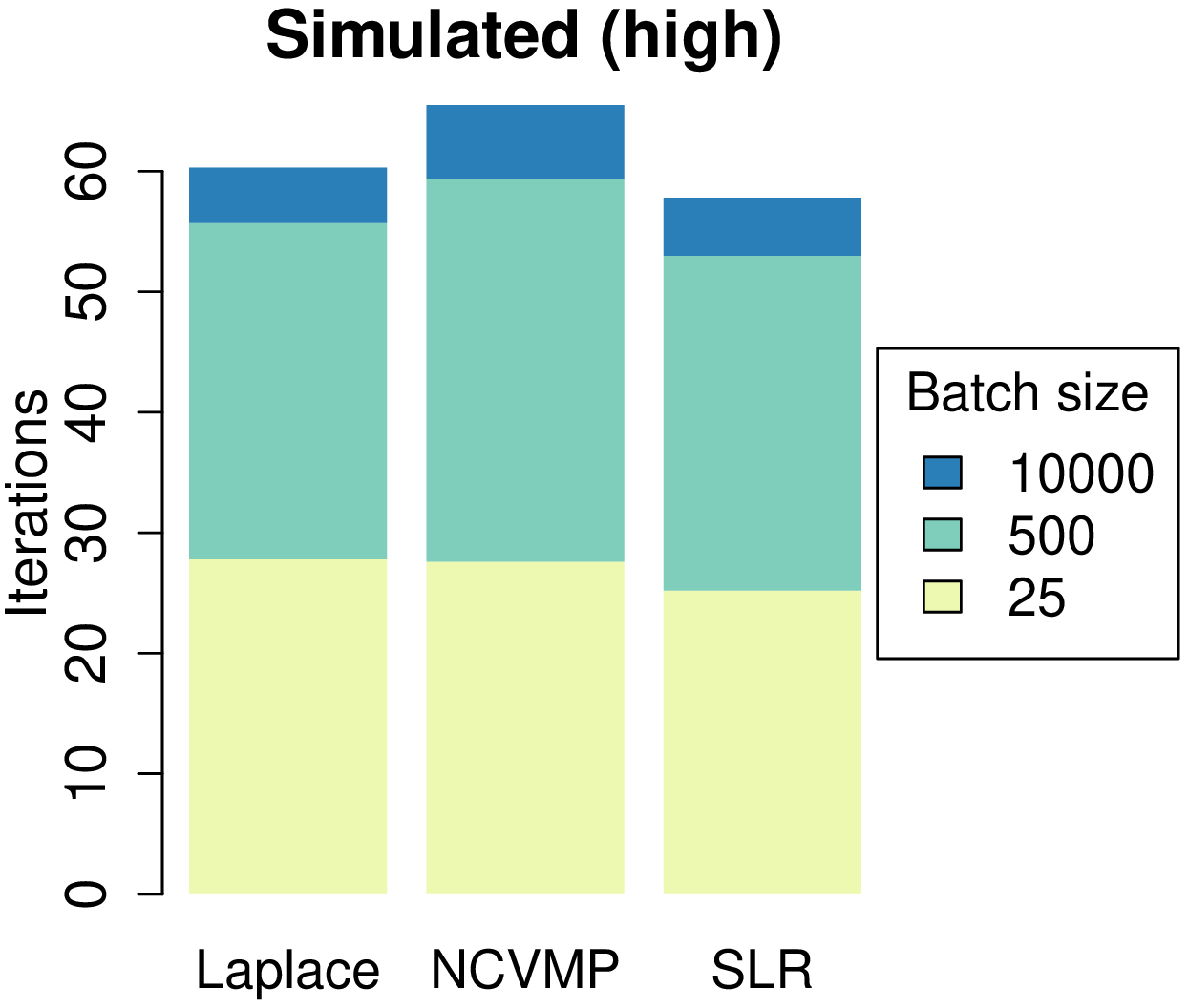} \includegraphics[width=0.238\textwidth]{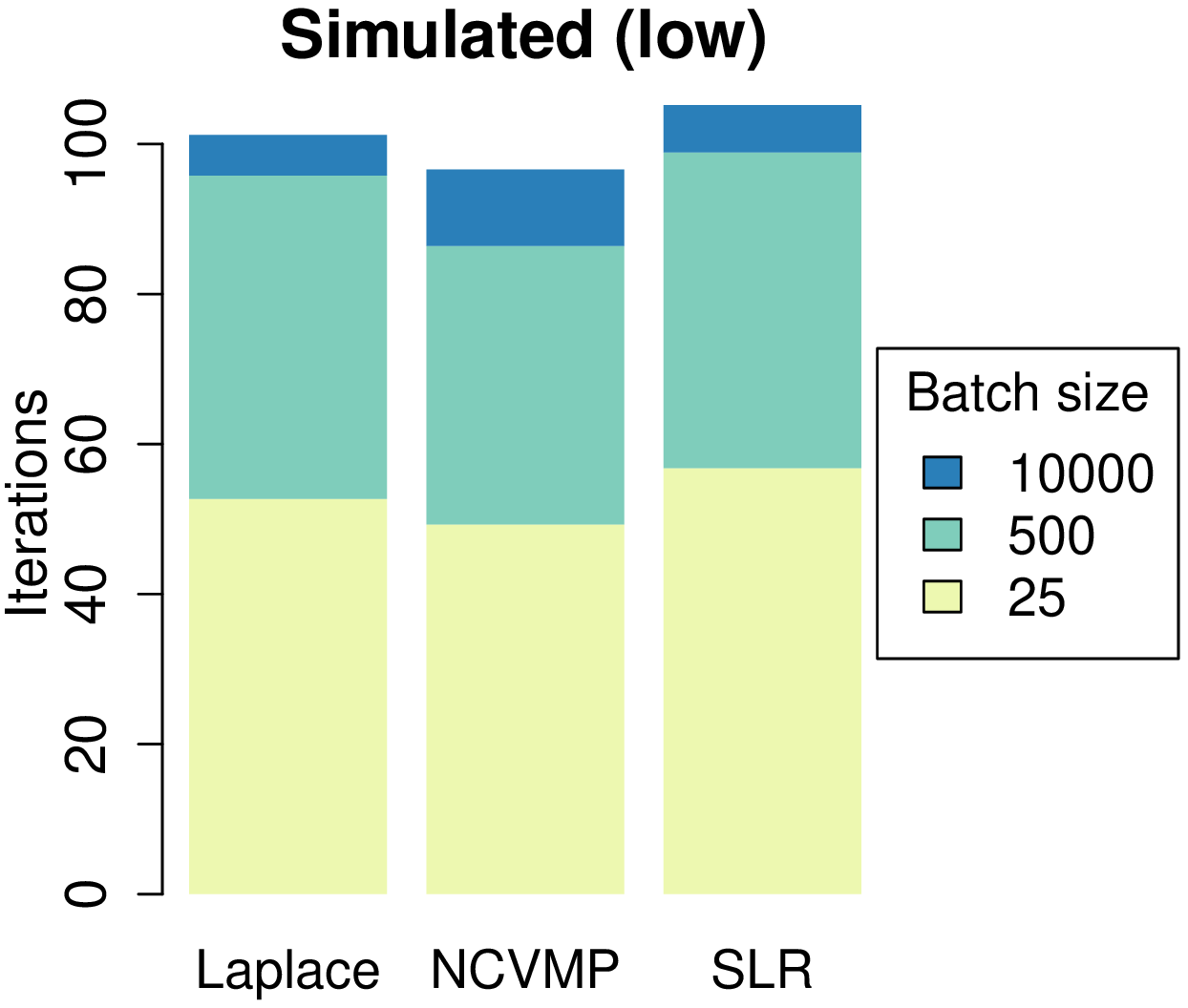}
\caption{\label{simulated}Barplots show the average number of iterations spent by Algorithm 2 (Laplace, NCVMP and SLR) at each minibatch size $|B|$ for the simulated data with high heterogeneity (left) and low heterogeneity (right).}
\end{figure}

\begin{table}[htb!]
	\centering \ra{1.1}
	\begin{tabular}{@{}llll@{}}
		\hline
		\multicolumn{4}{c}{High heterogeneity}    \\ \hline
		Methods & Algorithm 1 & Algorithm 2 & Reduction  \\ \hline
		Laplace   & 1008           & 470 (29)   & 53 \%\\
		NCVMP  & 432             & 311 (11)   & 28 \% \\
		SLR        & 1752 (5)      & 797 (51)   & 55 \% \\ \hline
		\multicolumn{4}{c}{Low heterogeneity}    \\ \hline
		Methods & Algorithm 1 & Algorithm 2 & Reduction  \\ \hline
		Laplace   & 1348            & 674 (51)     & 50 \%\\
		NCVMP  & 716              & 389 (42)     & 46 \% \\
		SLR        & 1752 (4)       & 1104 (110) & 37 \% \\ \hline
	\end{tabular}
	\caption{\label{simulated_runtime} Simulated data: CPU times (seconds) for Algorithms 1 and 2. Last column indicates percentage reduction in CPU times from using Algorithm 2 instead of 1. Standard deviation over repeated runs given in brackets.}
\end{table}
We run Algorithms 1 and 2 as well as MCMC on both datasets. Runtimes are given in Table \ref{simulated_runtime}. For Algorithm 2, we experimented with different values of $\kappa$ from 2 to 20. A larger $\kappa$ led to a greater reduction in computation time and the results in Table \ref{simulated_runtime} are for $\kappa = 20$. Generally, larger values of $\kappa$ seem to work better for larger datasets. In this example, the coefficients in $\beta_h$ were simulated independently and the variational algorithms in batch mode were able to move quickly to a point that is close to the optimum. Thus, the reduction in CPU times through the use of minibatches was limited. From Table \ref{simulated_runtime}, reductions in CPU times from using Algorithm 2 instead of 1 ranged from 28\% to 55\%. The average number of iterations Algorithm 2 spent at each minibatch size $|B|$ are shown in Figure \ref{simulated}. For MCMC, we run 4 chains, each with 10,000 iterations and a thinning factor of 2 was applied. 
\begin{table*}[htb!]
	\centering\ra{1.1}
	\begin{tabular}{@{}llcccccc@{}}
		\hline
		Heterogeneity & Methods & Min. & 1st Qu. & Median & Mean & 3rd Qu. & Max   \\ \hline
		\multirow{4}{*}{Low} & Laplace & 1.04 \% & 2.04 \% & 2.38 \% & 2.38 \% & 2.69 \% & 4.02 \% \\
		& NCVMP & 0.14 \% & 0.38 \% & 0.50 \% & 0.49 \% & 0.60 \% & 0.96 \% \\
		& SLR & 0.13 \% & 0.34 \% & 0.44 \% & 0.45 \% & 0.54 \% & 0.92 \%   \\
		& MCMC & 0.09 \% & 0.37 \% & 0.47 \% & 0.47 \% & 0.57 \% & 0.89 \% \\ \hline
		\multirow{4}{*}{High} & Laplace & 0.63 \% & 1.52 \% & 1.76 \% & 1.79 \% & 2.04 \% & 3.02 \% \\
		& NCVMP & 0.07 \% & 0.31 \% & 0.41 \% & 0.44 \% & 0.54 \% & 1.00 \% \\
		& SLR & 0.04 \% & 0.29 \% & 0.41 \% & 0.44 \% & 0.57 \% & 1.08 \%  \\
		& MCMC & 0.04 \% & 0.31 \% & 0.42 \% & 0.45 \% & 0.56 \% & 1.05 \% \\  \hline  
	\end{tabular}
	\caption{\label{simulated_TV} Simulated data: Summary of 500 TV distances between predictive choice probabilities computed using MCMC and Algorithm 1 (Laplace, NCVMP and SLR).}
\end{table*}

In this simulation study, the true predictive choice distribution can be computed. Table \ref{simulated_TV} provides a summary of the TV errors of MCMC and the variational methods from the true predictive choice distribution. The TV errors were computed at 500 attribute matrices $x_{\text{new}}$, the entries of which were generated randomly from $N(0,0.5^2)$. Table \ref{simulated_TV} shows that there is little difference in accuracy between NCVMP, SLR and MCMC while Laplace approximation did much worse than the rest. SLR did slightly better than NCVMP for the low heterogeneity case. 

\subsection{Project on faculty appointments}
We consider a subset of the data from The Project on Faculty Appointments, a study conducted at the Harvard Graduate School of Education  to examine the importance of different factors in job decisions \citep{Trower2002}. Survey respondents (consisting of faculty and doctoral candidates) were each presented with 16 pairs of job positions, and asked to select among either one of the two positions or neither, for each pair. The job positions varied along factors such as balance of work, chance of tenure or contract renewal, geographic location, department rating, institution rating, salary, tenure or non-tenure track and length of contract for non-tenured track. We consider $H = 1274 $ respondents with $T_h = 16$ for all $h$, $J=3$ and a total of $K=10$ covariates. These covariates are effect coded indicator variables for the factors described above, with between two to four levels.

\begin{figure}[htb!]
	\centering
	\includegraphics[width=0.28\textwidth]{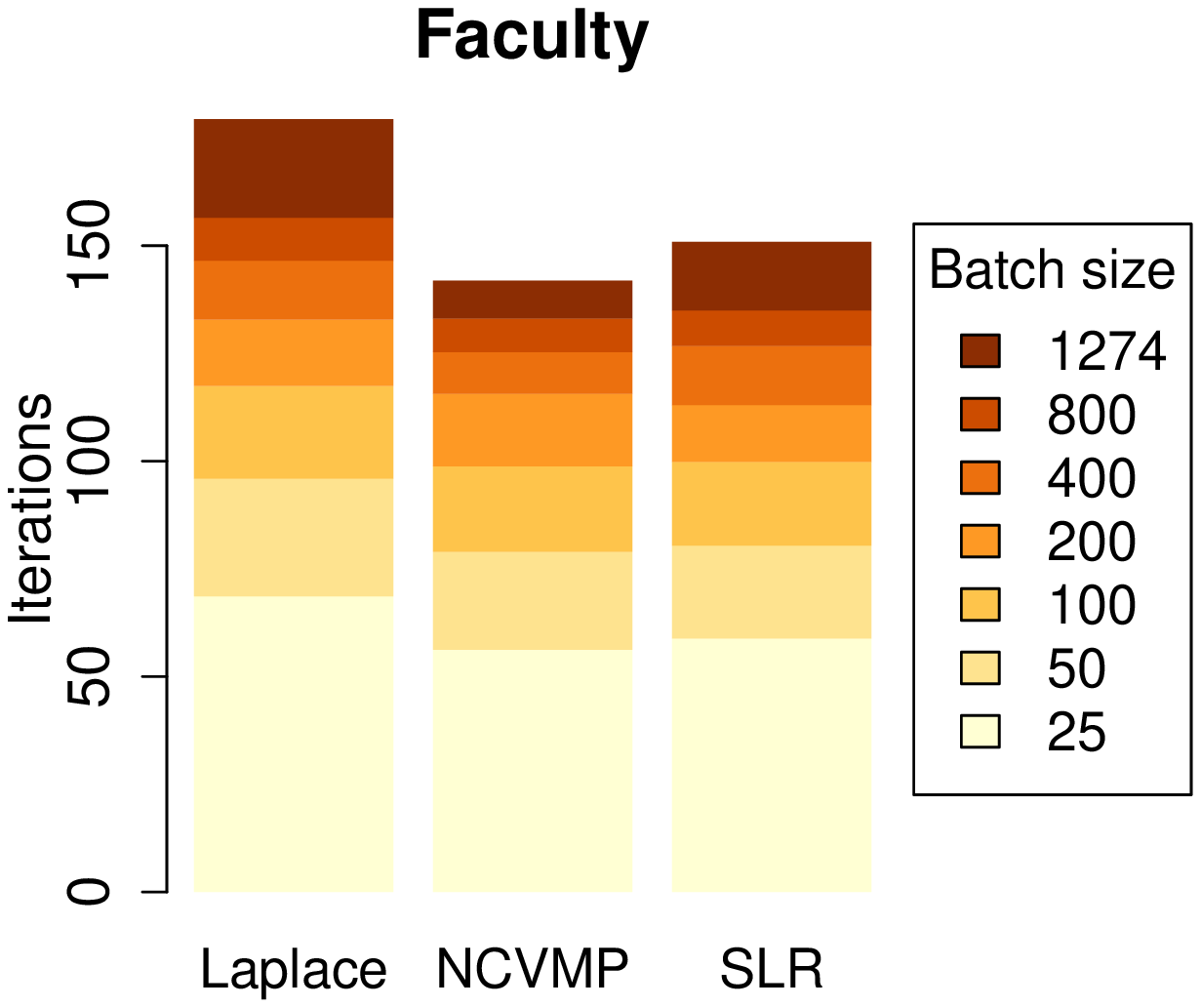}   
	\caption{\label{faculty} Barplot shows the average number of iterations spent by Algorithm 2 (Laplace, NCVMP, SLR) at each minibatch size $|B|$ for the faculty data.}
\end{figure}

Runtimes of Algorithms 1 and 2 and MCMC are given in Table \ref{faculty_runtime}. We used $\kappa = 2$ for Algorithm 2. Table \ref{simulated_runtime} indicates good reductions in CPU times of 54\% to 62\% when using Algorithm 2 instead of 1. The average number of iterations Algorithm 2 spent at each minibatch size $|B|$ is shown in Figure \ref{faculty}. For MCMC, we run 4 chains, each with 50000 iterations and a thinning factor of 10 was applied. For this dataset, parameters corresponding to several variables took a long time to reach convergence and there was also high correlation between draws. Hence, a large number of iterations and a very high thinning factor were used.
\begin{table}[htb!]
\centering\ra{1.1}
\begin{tabular}{@{}llll@{}}
\hline
Methods & Algorithm 1 & Algorithm 2 & Reduction  \\ \hline
Laplace & 707 & 325 (57) & 54 \%\\
NCVMP & 113 & 51 (5) & 55 \% \\
SLR & 714 (12) & 274 (39) & 62 \%   \\  \hline  
\end{tabular}
\caption{\label{faculty_runtime} Faculty example: CPU times (seconds) for Algorithms 1 and 2. Last column indicates percentage reduction in CPU times from using Algorithm 2 instead of 1. Standard deviation over repeated runs given in brackets.}
\end{table}

Table \ref{faculty_TV} shows a summary of the TV distances between the predictive choice distribution estimated using MCMC and the variational methods. The TV distances were computed at 1274 attribute matrices, obtained by randomly selecting one choice event covariate matrices $x_{ht}$ from each respondent. SLR produced results that were closest to that of MCMC with NCVMP close behind. Results from Laplace approximation were much further away from that of MCMC than NCVMP and SLR.
\begin{table*}[htb!]
	\centering
	\begin{tabular}{@{}llccccc@{}}
		\hline
		Methods & Min. & 1st Qu. & Median & Mean & 3rd Qu. & Max   \\ \hline
		Laplace vs. MCMC & 0.45 \% & 1.22 \% & 1.80 \% & 1.83 \% & 2.21 \% & 3.70 \% \\
		NCVMP vs. MCMC& 0.02 \% & 0.13 \% & 0.24 \% & 0.24 \% & 0.31 \% & 0.57 \% \\
		SLR vs. MCMC & 0.02 \% & 0.15 \% & 0.22 \% & 0.22 \% & 0.28 \% & 0.51 \%   \\
		\hline
	\end{tabular}
	\caption{\label{faculty_TV} Faculty example: Summary of 1274 TV distances between predictive choice probabilities computed using MCMC and Algorithm 1 (Laplace, NCVMP and SLR).}
\end{table*}

\subsection{Tuna data}
Discrete choice models play an important role in developing pricing strategies. In this example, we consider scanner data on canned tuna, which is available in the {\tt R} package {\tt Ecdat}. The total number of households is $H=3093$ and there are 13705 purchase records in total. The number of records for each household varies greatly, with $1 \leq T_h \leq 64$. At each purchase occasion, each household chooses one of $J=5$ brands of tuna: Starkist water, Chicken-of-the-Sea water, a store-specific private label (water), Starkist oil and Chicken-of-the-sea oil. We consider $K=2$ variables: price and an indicator variable for ``water''. More details about this dataset can be found in \cite{Kim1995}.

\begin{figure}[htb!]
	\centering
	\includegraphics[width=0.29\textwidth]{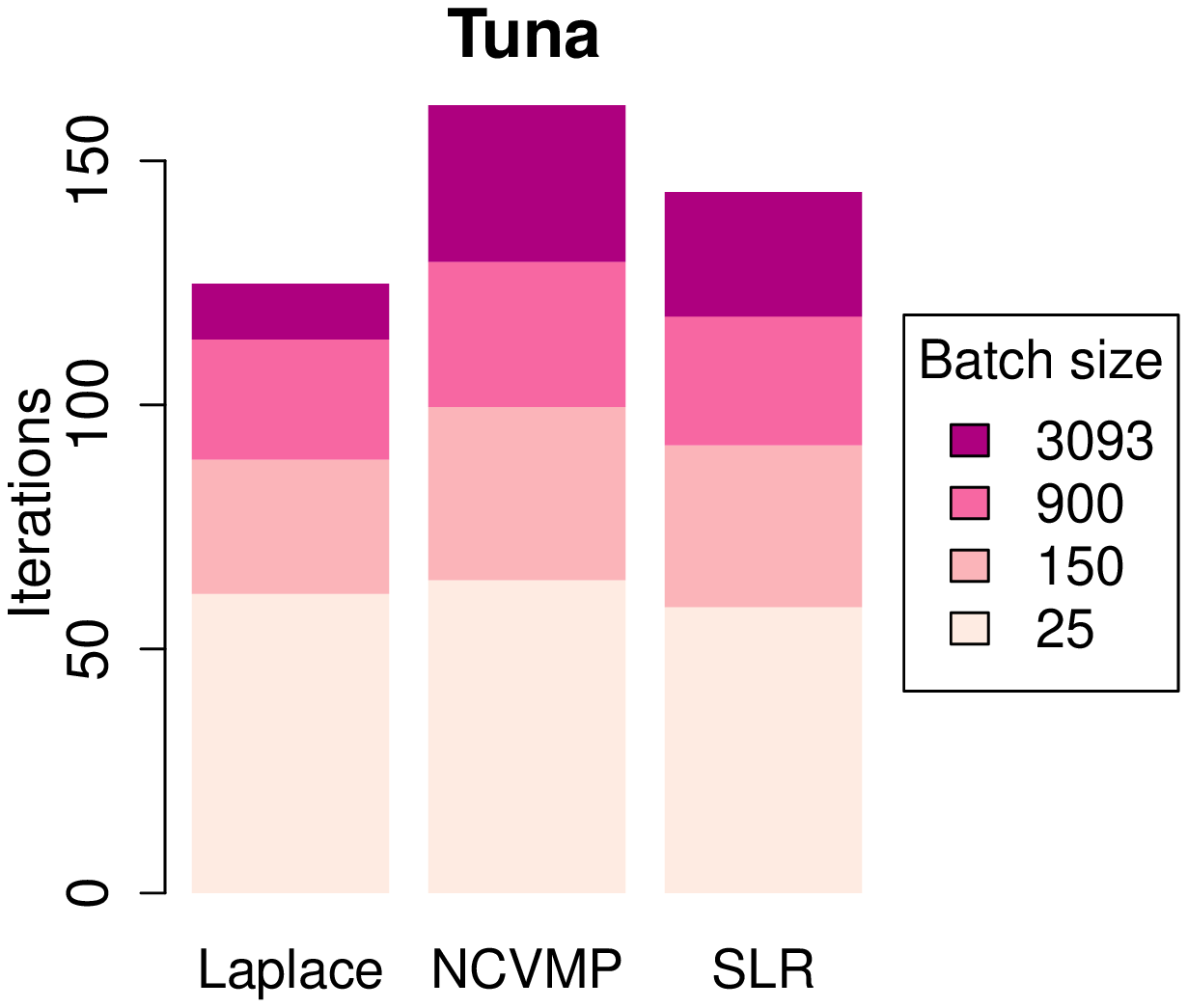}   
	\caption{\label{tuna} Barplot shows the average number of iterations spent by Algorithm 2 (Laplace, NCVMP, SLR) at each minibatch size $|B|$ for the tuna data.}
\end{figure}

We run Algorithms 1 and 2 and MCMC on this dataset. For Algorithm 2, we investigated values of $\kappa$ ranging from 2 to 10. Results are similar and Table \ref{tuna_runtime} shows the results for $\kappa=6$. There is a large amount of fluctuation between repeated runs of Algorithm 2. This is likely due to the  huge variation in number of purchase records for each household. Algorithm 2 reduced CPU times significantly for all three options: Laplace, NCVMP and SLR. The average number of iterations Algorithm 2 spent at each minibatch size $|B|$ is shown in Figure \ref{tuna}. For MCMC, we run 4 chains, each with 10,000 iterations and a thinning factor of 2 was applied.
\begin{table}[htb!]
	\centering\ra{1.1}
	\begin{tabular}{@{}llll@{}}
		\hline
		Methods & Algorithm 1 & Algorithm 2 & Reduction  \\ \hline
		Laplace  & 306         & 103 (33)   & 66 \%\\
		NCVMP & 100         & 64 (11)     & 36 \% \\
		SLR       & 770 (9)    & 409 (112) & 47 \%   \\ \hline  
	\end{tabular}
	\caption{\label{tuna_runtime} Tuna example: CPU times (seconds) for Algorithms 1 and 2. Last column indicates percentage reduction in CPU times from using Algorithm 2 instead of 1. Standard deviation over repeated runs given in brackets.}
\end{table}

\begin{table*}[htb!]
	\centering\ra{1.1}
	\begin{tabular}{@{}llccccc@{}}
		\hline
		Methods & Min. & 1st Qu. & Median & Mean & 3rd Qu. & Max   \\ \hline
		Laplace vs. MCMC & 0.99 \% & 2.27 \% & 3.34 \% & 3.38 \% & 4.28 \% & 5.61 \% \\
		NCVMP vs. MCMC& 0.14 \% & 0.73 \% & 0.92 \% & 0.96 \% & 1.08 \% & 2.48 \% \\
		SLR vs. MCMC & 0.06 \% & 0.46 \% & 0.76 \% & 0.77 \% & 0.95 \% & 1.52 \%   \\
		\hline
	\end{tabular}
	\caption{\label{tuna_TV} Tuna example: Summary of 1000 TV distances between predictive choice probabilities computed using MCMC and Algorithm 1 (Laplace, NCVMP and SLR).}
\end{table*}
Table \ref{tuna_TV} provides a summary of the TV distances between the predictive choice distribution estimated using MCMC and the variational methods. The TV distances were computed at 1000 attribute matrices, obtained by randomly selecting one choice event covariate matrix $x_{ht}$ from each of 1000 randomly chosen respondents. SLR produced results closest to that of MCMC, with NCVMP close behind. Laplace approximation does markedly worse than NCVMP and SLR. 

\subsection{Electricity data} \label{eg_electricity}
This dataset consists of $H=361$ residential electricity customers, who were each presented with 12 choice experiments. In each experiment, the respondent was asked to choose an electricity supplier out of $J=4$  alternatives after being presented with their attributes. Some respondents did not complete all 12 experiments and there were 4308 experiments with $8 \leq T_h \leq 12$. Attributes of the suppliers include price, contract length in years and whether the company was local or well-known. The price of the supplier was either a fixed price stated in cents per kWh, or a price plan offered in the form of a time-of-day rate or seasonal rates. There were $K=6$ variables in total. More details can be found in \cite{Hubert2001} and the data is available in the {\tt R} package {\tt mlogit}. 

\begin{table}[htb!]
	\centering\ra{1.1}
	\begin{tabular}{@{}llcc@{}}
		\hline
		Methods & Algorithm 1 & Algorithm 2 & Reduction  \\ \hline
		Laplace & 159 & 69 (4) & 57 \%\\
		NCVMP & Diverge & Diverge & -- \\
		SLR & 255 (6) & 157 (8) & 38 \%   \\ \hline  
	\end{tabular}
	\caption{\label{electricity_runtime} Electricity example: CPU times (seconds) for Algorithms 1 and 2. Last column indicates percentage reduction in CPU times from using Algorithm 2 instead of 1. Standard deviation over repeated runs given in brackets.}
\end{table}

\begin{figure}[htb]
	\centering
	\includegraphics[width=0.22\textwidth]{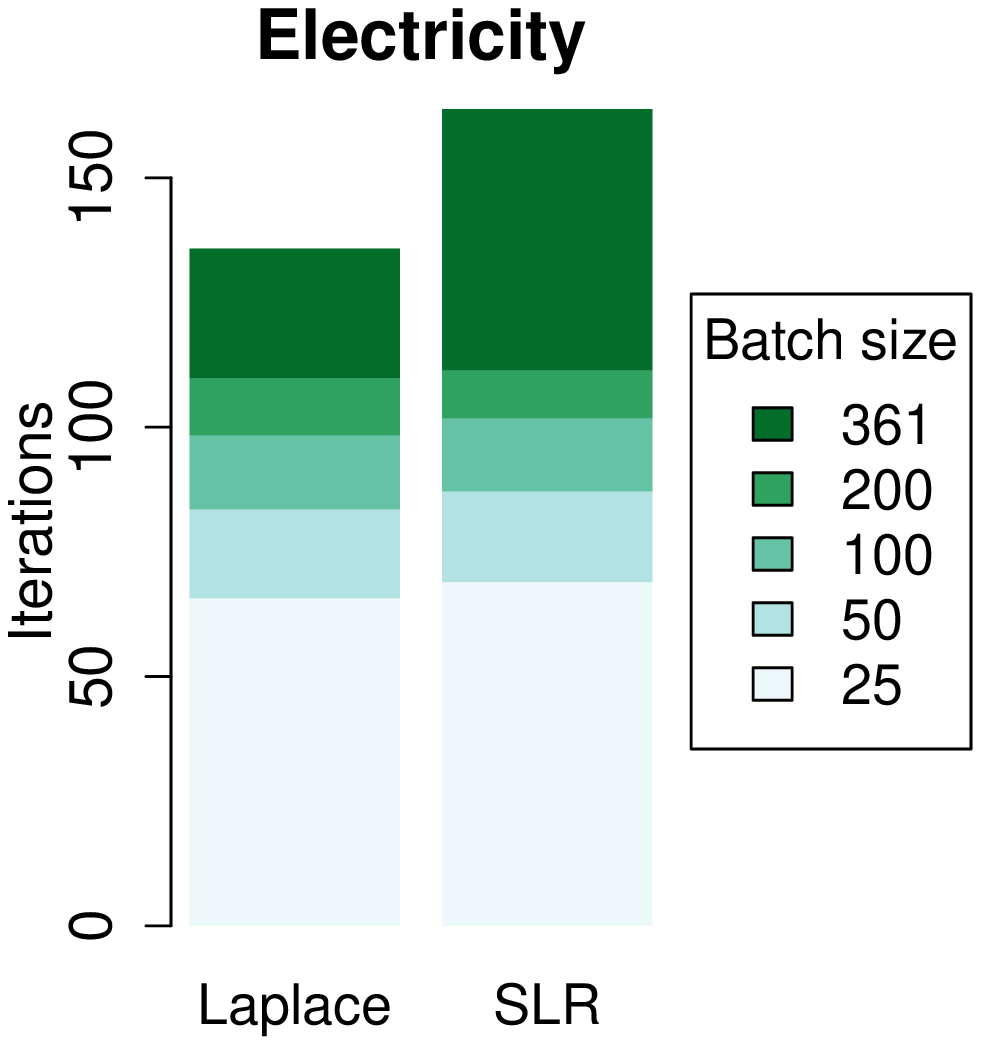}   
	\caption{\label{electricity} Barplot shows the average number of iterations spent by Algorithm 2 (Laplace, NCVMP, SLR) at each minibatch size $|B|$ for the electricity data.}
\end{figure}

Runtimes of Algorithms 1 and 2 and MCMC are given in Table \ref{electricity_runtime}. For this dataset, NCVMP failed to converge in both Algorithms 1 and 2. Failure to converge could be due to the fixed point iterations in NCVMP or the approximations in the delta method. We investigate the problem by optimizing $\mu_h$ and $\Sigma_h$ using the {\tt optim} function in {\tt R} instead of NCVMP. The algorithm failed to converge with {\tt optim} as well and thus the problem lies in the delta method approximation. There were no convergence issues with SLR. This example shows that SLR  can help to overcome convergence issues encountered in NCVMP and the delta method. However, SLR is slower and the delta approximation worked very well for all other examples. To avoid divergence issues and to speed up computations, one could run Algorithm 1 using NCVMP, monitor the lower bound, and switch to SLR when the lower bound fails to increase. While this dataset is relatively small, speedups can still be obtained using Algorithm 2. The results in Table \ref{electricity_TV} correspond to $\kappa=2$ for Algorithm 2. The average number of iterations Algorithm 2 spent at each minibatch size $|B|$ is shown in Figure \ref{electricity}. For MCMC, we run 4 chains, each with 10,000 iterations and a thinning factor of 2 was applied.

Table \ref{electricity_TV} provides a summary of the TV distances between the predictive choice distributions estimated using MCMC and the variational methods. The TV distances were computed at 1444 attribute matrices, obtained by randomly selecting four choice event covariate matrices $x_{ht}$ from each of the 361 respondents. The agreement between SLR and MCMC is very good. Discrepancy between Laplace approximation and MCMC is much more pronounced. 
\begin{table*}[htb!]
\centering\ra{1.1}
\begin{tabular}{@{}llccccc@{}}
\hline
Methods & Min. & 1st Qu. & Median & Mean & 3rd Qu. & Max   \\ \hline
Laplace vs. MCMC & 0.88 \% & 1.98 \% & 2.31 \% & 2.42 \% & 3.18 \% & 4.16 \% \\
SLR vs. MCMC & 0.15 \% & 0.36 \% & 0.41 \% & 0.43 \% & 0.50 \% & 0.73 \%   \\
\hline
\end{tabular}
\caption{\label{electricity_TV} Electricity example: Summary of 1444 TV distances between predictive choice probabilities computed using MCMC and Algorithm 1 (Laplace, NCVMP and SLR).}
\end{table*}

\section{Conclusion} \label{Conclusion}
In this paper, we have developed three different approaches for fitting MMNL models using variational Bayes: (1) Laplace approximation, (2) nonconjugate variational message passing with a delta method approximation and (3) stochastic linear regression. We also proposed a novel adaptive batch size strategy for implementing stochastic variational inference using these approaches. The performances of these variational methods were investigated for a wide range of datasets, both real and simulated. Across all examples, predictive choice distributions computed using stochastic linear regression were closest to that of MCMC, with nonconjugate variational message passing close behind. The discrepancy between Laplace approximation and MCMC is much more pronounced. In terms of stability, stochastic linear regression and Laplace approximation are very stable and we did not encounter any convergence issues in all our experiments. While nonconjugate variational message passing failed to converge in one of the examples, the failure is due to the delta method approximation rather than nonconjugate variational message passing itself. In the rest of the examples, nonconjugate variational message passing performed very well and converged in the shortest time. Stochastic variational inference further accelerates convergence for large scale data sets. With our adaptive batch size strategy, Algorithm 2 is nearly automatic and we recommend increasing $\kappa$ proportionately with the number of agents $H$. Significant speedups can be obtained using Algorithm 2 for datasets as small as a few hundreds. Variational methods provide an important alternative as well as complement to MCMC methods for fitting MMNL models, yielding high computational efficiency with competitive accuracy. Investigating how variational methods can be used for performing model selection with large-scale discrete choice datasets is another important area for future research. With the availability of large scale data sets in marketing and other applications, variational methods enable predictive inference to be obtained in a timely manner.

\begin{acknowledgements}
The author is grateful to Cathy Trower, Jordan Louviere and the Institute for Choice at the University of South Australia for sharing data from The Project on Faculty Appointments. The author also wish to thank Kenneth Train for sharing data from the vehicle choice stated preference experiments, David Nott for his comments on the manuscript, and the editor, associate editor and referees for their comments and suggestions for improving the manuscript. 
\end{acknowledgements}

\begin{appendices}
\renewcommand\thesection{Appendix \Alph{section}}

\section{Optimal densities of conjugate factors} \label{opt_den_deriv}
Optimal densities for $q(a)$, $q(\zeta)$ and $q(\Omega)$ can be derived using \eqref{optimal densities}. 
\begin{equation*}
\begin{aligned}
q(a) &\propto \exp E_{-a} \{ \log p(y,\theta) \} \\
       & \propto \prod_{k=1}^K \exp  \bigg \{ - \frac{\nu+K+2}{2}  \log a_k \\
       & \quad - \left( \nu E_{q(\Omega)}(\Omega_{kk}^{-1}) + \frac{1}{A_k^2}  \right)  \frac{1}{a_k} \bigg \}  \\
       & = \prod_{k=1}^K IG(b_k, c_k),
\end{aligned}
\end{equation*}
where 
\begin{gather*}
b_k = \frac{\nu+K}{2}, c_k = \nu E_{q(\Omega)}(\Omega_{kk}^{-1}) + \frac{1}{A_k^2}.
\end{gather*}

\begin{equation*}
\begin{aligned}
q(\zeta) &\propto \exp E_{-\zeta} \{ \log p(y,\theta) \} \\
            & \propto \exp - \frac{1}{2} \bigg [ \zeta^T \left\{ \Sigma_0^{-1} + H E_{q(\Omega)} (\Omega^{-1})  \right\}  \zeta \\
            & \quad -2\zeta^T \bigg\{  \Sigma_0^{-1}\mu_0  +  E_{q(\Omega)} (\Omega^{-1}) \sum_{h=1}^H E_{q(\beta_h)} (\beta_h)  \bigg \} \bigg] \\
            &= N(\mu_\zeta, \Sigma_\zeta),
\end{aligned}
\end{equation*}
where
\begin{gather*}
\Sigma_\zeta = \left\{ \Sigma_0^{-1} + H E_{q(\Omega)} (\Omega^{-1})  \right\} ^{-1}, \\
\mu_\zeta =  \Sigma_\zeta  \left\{ \Sigma_0^{-1}\mu_0  +  E_{q(\Omega)} (\Omega^{-1}) \sum_{h=1}^H \mu_h  \right\}.
\end{gather*}

\begin{equation*}
\begin{aligned}
& q(\Omega) \\
& \propto \exp E_{-\Omega} \{ \log p(y,\theta) \} \\
& \propto \exp E_{-\Omega} \bigg\{ - \frac{H+\nu+2K}{2} \log |\Omega| \\
&\quad  - \frac{1}{2} \sum_{h=1}^H (\beta_h -\zeta)^T \Omega^{-1} (\beta_h -\zeta) - \nu \sum_{k=1}^K \frac{\Omega_{kk}^{-1}}{a_k}   \bigg\} \\
& \propto \exp \bigg[ - \frac{H+\nu+2K}{2} \log |\Omega| - \frac{1}{2} \text{tr}\bigg\{ \Omega^{-1} \\
& \quad E_{-\Omega}\left( \sum_{h=1}^H (\beta_h -\zeta)(\beta_h -\zeta)^T + 2\nu \text{diag}\left( \frac{1}{a} \right)  \right)  \bigg\}  \bigg] \\
& = IW (\omega, \Upsilon)   
\end{aligned}
\end{equation*}
where 
\begin{multline*}
\omega = H+\nu+K-1, \\
\Upsilon = \sum_{h=1}^H \{(\mu_h -\mu_\zeta) (\mu_h -\mu_\zeta)^T + \Sigma_h+ \Sigma_\zeta \}+ 2\nu \text{diag}  \left(\frac{b}{c} \right).
\end{multline*}

\section{Variational lower bound}
\begin{equation*}
\begin{aligned}
&\log q(\theta) \\
&= \sum_{h=1}^H \log q(\beta_h)  + \log q(\zeta) + \log q(\Omega) + \sum_{k=1}^K \log q(a_k) \\
&= - \frac{1}{2}  \sum_{h=1}^H (\beta_h -\mu_h)^T \Sigma_h^{-1} (\beta_h -\mu_h) - \frac{1}{2}  \sum_{h=1}^H \log |\Sigma_h|  \\
& \quad  -\frac{(H+1)K}{2} \log (2\pi) - \frac{1}{2} (\zeta -\mu_\zeta)^T \Sigma_\zeta^{-1} (\zeta -\mu_\zeta)  \\
&\quad  - \frac{1}{2} \log |\Sigma_\zeta| -\frac{\omega K}{2} \log 2 - \frac{K(K-1)}{4} \log \pi \\
&\quad  - \sum_{k=1}^K \log \Gamma \left( \frac{\omega+1-k}{2} \right)+ \frac{\omega}{2} \log |\Upsilon| - \frac{1}{2} \text{tr}(\Upsilon \Omega^{-1})  \\
& \quad + \sum_{k=1}^K \bigg( b_k \log c_k - \log \Gamma(b_k) - (b_k + 1) \log a_k -\frac{c_k}{a_k} \bigg) \\
& \quad - \frac{\omega+K+1}{2} \log |\Omega|. \\
\end{aligned}
\end{equation*}

\begin{equation*}
\begin{aligned}
&\log p(y, \theta) \\
&= \sum_{h=1}^H \sum_{t=1}^{T_h} \log p(y_{ht}|x_{ht}, \beta_h)  + \sum_{h=1}^H \log p(\beta_h|\zeta, \Omega) \\
&\quad + \log p(\zeta| \mu_0, \Sigma_0) + \log p(\Omega|\nu,a) + \sum_{k=1}^K \log p(a_k|A_k) \\
&=  \sum_{h=1}^H \sum_{t=1}^{T_h} \left[ y_{ht}^T x_{ht} \beta_h-  \log \left\{ \sum_{j=1}^J \exp (x_{htj}^T \beta_h) \right\} \right]  \\
& \quad + \sum_{h=1}^H 
 \left\{- \frac{1}{2} \log |\Omega| - \frac{1}{2} (\beta_h -\zeta)^T \Omega^{-1} (\beta_h -\zeta)  \right\}  \\
& \quad  -\frac{(H+1)K}{2} \log (2\pi) - \frac{1}{2} (\zeta -\mu_0)^T \Sigma_0^{-1} (\zeta -\mu_0) \\
& \quad +\frac{(\nu+K-1)K}{2} \log \nu - \sum_{k=1}^K \log \Gamma \left( \frac{\nu+K-k}{2} \right) \\
& \quad - \frac{\nu+K-1}{2}\sum_{k=1}^K \log a_k  - \frac{1}{2} \log |\Sigma_0|- \nu \sum_{k=1}^K \frac{\Omega_{kk}^{-1}}{a_k}\\
& \quad + \sum_{k=1}^K\left\{ \log A_k - \log \Gamma\left(\frac{1}{2}\right) -\frac{3}{2} \log a_k -\frac{1}{A_k^2 a_k} \right\} \\
& \quad - \frac{\nu+2K}{2} \log |\Omega|  - \frac{K(K-1)}{4} \log \pi.
\end{aligned}
\end{equation*}

\begin{align*}
&E_q \{\log q(\theta)\} \\
&= -\frac{(H+1)K}{2} \{\log (2\pi) +1\} - \frac{1}{2} \sum_{h=1}^H \log |\Sigma_h| \\
&\quad - \frac{1}{2} \log |\Sigma_\zeta|  -\frac{\omega K}{2} (\log 2+1) - \frac{K(K-1)}{4} \log \pi \\
&\quad - \sum_{k=1}^K \log \Gamma \left( \frac{\omega+1-k}{2} \right)+ \frac{K+1}{2} \log |\Upsilon| \\
& \quad + \frac{\omega+K+1}{2} \left\{\sum_{k=1}^K \psi\left( \frac{\omega-k+1}{2} \right) + K \log 2  \right\} \\
& \quad + \sum_{k=1}^K [ (b_k + 1) \psi(b_k)- \log c_k - \log \Gamma(b_k) -b_k ].
\end{align*}

\begin{align*}
& E_q  \{\log p(y, \theta)\} \\
&= \sum_{h=1}^H E_{q(\beta_h)} \{f(\beta_h)\} - \frac{1}{2} (\mu_\zeta -\mu_0)^T \Sigma_0^{-1} (\mu_\zeta -\mu_0)  \\
& \quad - \frac{\nu+K+2}{2} \sum_{k=1}^K \{ \log c_k - \psi(b_k) \}  -\frac{\omega H}{2} \text{tr}(\Sigma_\zeta \Upsilon^{-1})\\
& \quad- \frac{\nu+2K+H}{2} \Big\{ \log |\Upsilon|-\sum_{k=1}^K \psi\left( \frac{\omega-k+1}{2} \right) \\
& \quad  -K \log 2  \Big\} - \frac{1}{2} \log |\Sigma_0| -\frac{(H+1)K}{2} \log (2\pi) \\
& \quad  - \frac{1}{2} \text{tr} (\Sigma_0^{-1} \Sigma_\zeta)  - \frac{K(K-1)}{4} \log \pi - K \log \Gamma\left(\frac{1}{2}\right) \\
& \quad - \sum_{k=1}^K \log \Gamma \left( \frac{\nu+K-k}{2} \right) + \frac{(\nu+K-1)K}{2} \log \nu \\
& \quad - \sum_{k=1}^K \left(  \nu \omega \Upsilon_{kk}^{-1} + \frac{1}{A_k^2} \right)  \frac{b_k}{c_k} -\sum_{k=1}^K \log A_k,
\end{align*}
where $f(\beta_h)$ is as defined in \eqref{fbetah}. 

Noting that the updates $b_k = \frac{\nu+K}{2}$ for $k=1,\dots,K$ and $\omega=H+\nu+K-1$ are deterministic, the variational lower bound $\mathcal{L}$ can be simplified as
\begin{equation*}
\begin{aligned}
\mathcal{L} &= \sum_{h=1}^H E_q \{f(\beta_h)\}-\frac{\omega H}{2} \text{tr}(\Sigma_\zeta \Upsilon^{-1}) - \frac{\omega}{2} \log |\Upsilon|\\
& \quad - \frac{1}{2} (\mu_\zeta -\mu_0)^T \Sigma_0^{-1} (\mu_\zeta -\mu_0)  - \frac{1}{2} \text{tr} (\Sigma_0^{-1} \Sigma_\zeta) \\
& \quad   - \sum_{k=1}^K \left(  \nu \omega \Upsilon_{kk}^{-1} + \frac{1}{A_k^2} \right)  \frac{b_k}{c_k} - \frac{1}{2} \log |\Sigma_0|  \\
& \quad + \frac{1}{2} \sum_{h=1}^H \log |\Sigma_h| + \frac{1}{2} \log |\Sigma_\zeta| - \sum_{k=1}^K b_k \log c_k \\
&\quad + \frac{(H+1+\omega + \omega \log 2)K}{2} + \frac{(\nu+K-1)K}{2} \log \nu \\
&\quad + \sum_{k=1}^K \bigg\{\log \Gamma \left( \frac{\omega+1-k}{2} \right) - \log \Gamma \left( \frac{\nu+K-k}{2} \right) \\
& \quad+ \log \Gamma(b_k) + b_k - \log A_k \bigg\} - K \log \Gamma\left(\frac{1}{2}\right).
\end{aligned}
\end{equation*}

\subsection{Variational objective for Laplace approximation} \label{appendix_Laplace}
From \eqref{Laplace_approx}, 
\begin{equation*}
f(\beta_h) \approx f(\hat{\beta}_h) + \frac{1}{2} (\beta_h-\hat{\beta}_h)^T H(\hat{\beta}_h)  (\beta_h-\hat{\beta}_h).
\end{equation*}
Since $q(\beta_h)= N(\mu_h, \Sigma_h)$ with $\mu_h=\hat{\beta}$ and $\Sigma_h=-H(\hat{\beta}_h)^{-1}$,
\begin{align*}
E_q \{f(\beta_h)\} &\approx f(\hat{\beta}_h) + \frac{1}{2} (\mu_h-\hat{\beta}_h)^T H(\hat{\beta}_h)  (\mu_h-\hat{\beta}_h) \\
& \quad + \frac{1}{2} \text{tr}(H(\hat{\beta}_h)\,\Sigma_h) \\
& = f(\hat{\beta}_h) -\frac{K}{2}.
\end{align*}
Using this approximation in the variational lower bound gives us an estimate of $\mathcal{L}$.

\subsection{Variational objective for delta method} \label{appendix_delta}
From \eqref{delta_approx}, $E_q \{f(\beta_h)\}$ is approximately given by 
\begin{multline*}
\sum_{t=1}^{T_h} \Bigg[  y_{ht}^T x_{ht} \mu_h 
- \log \left\{ \sum_{j=1}^J \exp \left(x_{htj}^T \mu_h \right)  \right\} \\
 - \frac{1}{2} \text{tr} \left\{ 
x_{ht}^T \left(  \text{diag}(\rho_{ht}) - \rho_{ht} \rho_{ht}^T \right)   
x_{ht} \Sigma_h  \right\} \Bigg] \\
- \frac{ \omega}{2} (\mu_h -\mu_\zeta)^T \Upsilon^{-1} (\mu_h -\mu_\zeta) - \frac{ \omega}{2} \text{tr} (\Sigma_h \Upsilon^{-1}).
\end{multline*}
Using vector differential calculus, it can be shown that 
\begin{multline*}
\frac{\partial E_q {\log p(y,\theta)}}{\partial \text{vec} (\Sigma_h)} =  \\
 - \frac{1}{2} \text{vec} \bigg\{ \sum_{t=1}^{T_h} x_{ht}^T ( \text{diag}(\rho_{ht}) - \rho_{ht} \rho_{ht}^T )x_{ht} + \omega \Upsilon^{-1} \bigg\},
\end{multline*}
\begin{multline*}
\frac{\partial E_q {\log p(y,\theta)}}{\partial \mu_h} = \sum_{t=1}^{T_h}  x_{ht}^T \bigg[ y_{ht}-\rho_{ht} + ( \text{diag}(\rho_{ht}) - \rho_{ht} \rho_{ht}^T) \\
 \times \Big\{ x_{ht} \Sigma_h x_{ht}^T \rho_{ht} - \frac{1}{2} \text{diag}(x_{ht} \Sigma_h x_{ht}^T) \Big\} \bigg]- \omega\Upsilon^{-1} (\mu_h-\mu_\zeta).
\end{multline*}

\section{Gradient and natural gradient of variational lower bound} \label{appendix_lb_grad}

The gradient of the variational lower bound with respect to $\lambda_i$ is
\begin{equation*} \label{NCVMPgrad}
\nabla_{\lambda_i} \mathcal{L} = \nabla_{\lambda_i} \int q(\theta) \log p(y,\theta) d\theta - \nabla_{\lambda_i} \int q(\theta) \log q(\theta) d\theta
\end{equation*}
The first term $\nabla_{\lambda_i} \int q(\theta) \log p(y,\theta) \;d\theta$ can be written as $\nabla_{\lambda_i}\, E_q\{\log p(y,\theta)\}$. Alternatively,
\begin{equation*}
\begin{aligned}
&\nabla_{\lambda_i} \int q(\theta) \log p(y,\theta) \;d\theta \\
&=\nabla_{\lambda_i}   \int  \prod_{i=1}^m q_i(\theta_i) \log p(y,\theta) \,d\theta  \\
& = \nabla_{\lambda_i}   \int q_i(\theta_i)  E_{-q_i} \{ \log p(y,\theta)\}  \,d\theta_i   \\
&= \int q_i(\theta_i) \{t_i(\theta_i)- \nabla_{\lambda_i} h_i(\lambda_i)\} E_{-q_i} \{ \log p(y,\theta)\}  \,d\theta_i  \\
&= E_{q_i} \left\{ t_i(\theta_i) E_{-q_i} \{ \log p(y,\theta)\} \right\} \\
& \quad - \nabla_{\lambda_i} h_i(\lambda_i) E_{q_i} \left\{ E_{-q_i} \{ \log p(y,\theta)\}  \right\} \\
&= \text{Cov}_{q_i} \left\{ t_i(\theta_i), E_{-q_i} \{ \log p(y,\theta)\} \right\},
\end{aligned}
\end{equation*}
since $E_{q_i} \{ t_i(\theta_i) \}  = \nabla_{\lambda_i} h_i(\lambda_i)$. The second term is
\begin{equation*}
\begin{aligned}
&\nabla_{\lambda_i} \int q(\theta) \log q(\theta) \;d\theta \\
& = \nabla_{\lambda_i} \int \prod_{i=1}^m q_i(\theta_i) \sum_{i=1}^m  \log q_i(\theta_i)   \;d\theta  \\
& = \nabla_{\lambda_i} \int q_i(\theta_i) \log q_i(\theta_i)  \,d\theta_i \\
& = \int q_i(\theta_i) \{t_i(\theta_i)- \nabla_{\lambda_i} h_i(\lambda_i)\}\{t_i(\theta_i)^T\lambda - h_i(\lambda_i)\} \,d\theta_i  \\
& \qquad + \int q_i(\theta_i)  \{t_i(\theta_i)- \nabla_{\lambda_i} h_i(\lambda_i)\}  \,d\theta_i \\
&= \left[ E_{q_i} \{ t_i(\theta_i)t_i(\theta_i)^T\} - \nabla_{\lambda_i} h_i(\lambda_i) E_{q_i}\{ t_i(\theta_i) \} \right] \lambda_i + 0 \\
&= \text{Cov}_{q_i} [t_i(\theta_i)] \lambda_i.
\end{aligned}
\end{equation*}
When $q_i$ is a member of the exponential family, the Fisher information matrix is given by
\begin{align*}
& E_{q_i} \left[ \left\{ \nabla_{\lambda_i} \log q_i(\theta_i) \right\} \left\{ \nabla_{\lambda_i} \log q_i(\theta_i) \right\} ^T\right] \\
&= E_{q_i} \left[ \left\{  t_i(\theta_i)-\nabla_{\lambda_i} h_i(\lambda_i) \right\} \left\{ t_i(\theta_i)-\nabla_{\lambda_i}h_i(\lambda_i) \right\} ^T\right]  \\
&= \text{Cov}_{q_i} [t_i(\theta_i)].
\end{align*}

\section{Stochastic linear regression multivariate Gaussian updates} \label{SLRmG}
When $q_i = N(\mu_i, \Sigma_i)$, the updates of $\mu_i$ and $\Sigma_i$ from nonconjugate variational message passing \eqref{NCVMP Gaussian} are 
\begin{multline*} 
\Sigma_i \leftarrow -\left[2 \text{vec}^{-1}\left( \nabla\negthinspace_{\text{vec}(\Sigma_i)} E_q\{\log p(y,\theta)\} \right)\right]^{-1} \;\; \text{and} \\ \mu_i \leftarrow \mu_i + \Sigma_i \;  \nabla\negthinspace_{\mu_i} E_q\{\log p(y,\theta)\}.
\end{multline*}
Using the identities in \eqref{Opper}, we have 
\begin{equation*}
\begin{aligned}
\Sigma_i^{-1} &= -2 \text{vec}^{-1}\left( \nabla\negthinspace_{\text{vec}(\Sigma_i)} E_{q_i} [E_{-q_i}\{\log p(y,\theta)\}] \right)\\
&=  -2 \text{vec}^{-1}\left( \frac{1}{2} \text{vec} \left\{ E_{q_i} [ \nabla_{\theta_i}^2 E_{-q_i}\{\log p(y,\theta)\}] \right\} \right)\\
&=  - E_{q_i} [ \nabla_{\theta_i}^2 E_{-q_i}\{\log p(y,\theta)\}] \\
\end{aligned}
\end{equation*}
and 
\begin{equation*}
\begin{aligned}
\mu_i &=E_{q_i}(\theta_i) + \Sigma_i \; \nabla\negthinspace_{\mu_i} E_{q_i} [E_{-q_i}\{\log p(y,\theta)\}]\\
 &=E_{q_i}(\theta_i) + \Sigma_i \; E_{q_i} [\nabla_{\theta_i} E_{-q_i}\{\log p(y,\theta)\}].
\end{aligned}
\end{equation*}
Hence, we have derived the updates in \eqref{SLRt} from nonconjugate variational message passing updates for multivariate Gaussian.

We provide a proof below for the identities in \eqref{Opper}, which were first derived by \cite{Minka2001} and \cite{Opper2009}. For simplicity, we omit the subscript $i$. For $q(\theta) = N(\mu,\Sigma)$ and any real function $V(\theta)$, we show that 
\begin{gather}
\nabla\negthinspace_{\mu} E_{q} \{ V(\theta)\} = E_{q}\{ \nabla_{\theta} V(\theta) \} \; \text{and}  \label{mn} \\ 
\nabla\negthinspace_{\text{vec}(\Sigma)} E_{q} \{ V(\theta)\} = \frac{1}{2}\text{vec} [E_{q}\{\nabla_{\theta}^2 V(\theta) \}]. \label{vr}
\end{gather}

\begin{proof}
\begin{multline*}
\text{LHS of \eqref{mn}}: \nabla\negthinspace_{\mu} E_{q} \{ V(\theta)\} =  \int \nabla\negthinspace_{\mu}q(\theta)\; V(\theta) \;d\theta \\
=  \int q(\theta)  \Sigma^{-1} (\theta - \mu) V(\theta) \;d\theta
\end{multline*}
Using integration by parts, the $i$th element of $E_{q}\{ \nabla_{\theta} V(\theta) \}$ is 
\begin{multline*}
\int q(\theta) \frac{\partial V(\theta)}{\partial \theta_i} d\theta = \int \left\{ 0 - \int V(\theta) \frac{\partial q(\theta)}{\partial \theta_i} d\theta_i  \right\}  d\theta_{-i}\\
 =  - \int V(\theta) \frac{\partial q(\theta)}{\partial \theta_i}  d\theta.
\end{multline*}
Therefore
\begin{multline*}
\text{RHS of \eqref{mn}}: E_{q}\{ \nabla_{\theta} V(\theta) \} = - \int V(\theta) \nabla_\theta q(\theta) d\theta \\
= - \int V(\theta)  \{-q(\theta) \Sigma^{-1} (\theta - \mu) \} \;d\theta. 
\end{multline*}
\end{proof}

\begin{proof}
\begin{equation*}
\begin{aligned}
&\text{LHS of \eqref{vr}}: \nabla\negthinspace_{\text{vec}(\Sigma)} E_{q} \{ V(\theta)\} \\
&=  \int \nabla\negthinspace_{\text{vec}(\Sigma)}\; q(\theta) V(\theta) \;d\theta \\
&= \frac{1}{2} \int q(\theta) \big[ (\Sigma^{-1} \otimes \Sigma^{-1}) \text{vec} \{(\theta-\mu)(\theta-\mu)^T\} \\
& \quad - \text{vec}(\Sigma^{-1}) \big] V(\theta) \;d\theta.
\end{aligned}
\end{equation*}
The $(i,j)$th element of $E_{q}\{\nabla_{\theta}^2 V(\theta) \}$ is 
\begin{equation*}
\begin{aligned}
\int q(\theta) \frac{\partial^2 V(\theta)}{\partial \theta_i \partial \theta_j} \;d\theta &= \int \left\{ 0 - \int \frac{\partial q(\theta)}{\partial \theta_j} \frac{\partial V(\theta)}{\partial \theta_i} d\theta_j  \right\}  d\theta_{-j} \\
&= -  \int  \frac{\partial q(\theta)}{\partial \theta_j} \frac{\partial V(\theta)}{\partial \theta_i}  d\theta \\
&= - \int \left\{ 0 - \int \frac{\partial^2 q(\theta)}{\partial \theta_i \partial \theta_j} V(\theta) d\theta_i  \right\}  d\theta_{-i} \\
&= \int \frac{\partial^2 q(\theta)}{\partial \theta_i \partial \theta_j} V(\theta)  \;d\theta,
\end{aligned}
\end{equation*}
applying integration by parts twice.
Therefore 
\begin{equation*}
\begin{aligned}
&\text{RHS of \eqref{vr}}: \frac{1}{2}\text{vec} [E_{q}\{\nabla_{\theta}^2 V(\theta) \}] \\
&= \frac{1}{2}\text{vec}\left[ \int \nabla_\theta^2 q(\theta) \; V(\theta)  \;d\theta \right] \\ 
&= \frac{1}{2}\text{vec}\bigg[ \int  q(\theta) \left\{ \Sigma^{-1} (\theta - \mu)(\theta - \mu)^T \Sigma^{-1} - \Sigma^{-1}  \right\}  \\
&\quad V(\theta)  \;d\theta \bigg]  \\
&= \frac{1}{2} \int q(\theta) [ (\Sigma^{-1} \otimes \Sigma^{-1}) \text{vec} \{(\theta-\mu)(\theta-\mu)^T\} \\
&\quad - \text{vec}(\Sigma^{-1}) ] V(\theta) \;d\theta.
\end{aligned}
\end{equation*}
\end{proof}

\end{appendices}


\begin{thebibliography}{References}

\bibitem[Amari(1998)]{Amari1998}
Amari, S. (1998).
Natural gradient works efficiently in learning.
{\it Neural Computation}, 10, 251--276.

\bibitem[Attias(1999)]{Attias1999}
Attias, H. (1999).
Inferring parameters and structure of latent variable models by variational Bayes.
In {\it Proceedings of the 15th Conference on Uncertainty in Artificial Intelligence} (eds. Laskey, K. and Prade, H.), 21--30.
San Francisco, CA: Morgan Kaufmann.

\bibitem[Ben-Akiva and Lerman(1985)]{Ben-Akiva1985}
Ben-Akiva, M. E. and Lerman, S. R. (1985). 
{\it Discrete choice analysis: theory and application to travel demand} (Vol. 9). 
USA: MIT press.

\bibitem[Bhat(1998)]{Bhat1998}
Bhat, C. (1998).
Accommodating variations in responsiveness to level-of-service variables in travel mode choice models. 
{\it Transportation Research A}, 32, 455--507.

\bibitem[Boyles {\it et al.}(2011)]{Boyles2011}
Boyles, L., Korattikara, A., Ramanan, D. and Welling, M. (2011).
Statistical tests for optimization efficiency.
In {\it Advances in Neural Information Processing Systems 24} (eds. Shawe-Taylor, J., Zemel, R. S., Bartlett, P. L., Pereira, F. and Weinberger, K.Q.), 2196--2204. 
NY USA: Curran Associates, Inc.

\bibitem[Braun and McAuliffe(2010)]{Braun2010}
Braun, M. and McAuliffe, J. (2010).
Variational inference for large-scale models of discrete choice.
{\it Journal of the American Statistical Association}, 105, 324--335.

\bibitem[Brownstone and Train(1999)]{Brownstone1999}
Brownstone, D. and Train, K.  (1999).
Forecasting new product penetration with flexible substitution patterns.
{\it Journal of Econometrics}, 89, 109--129.

\bibitem[Bickel and Doksum(2007)]{Bickel2007}
Bickel, P. J. and Doksum, K. A. (2007).
{\it Mathematical Statistics: Basic Ideas and Selected Topics} (2nd ed.), Vol 1., Upper Saddle River, NJ: Pearson Prentice Hall.

\bibitem[Gaivoronski(1988)]{Gaivoronski1988}
Gaivoronski, A. (1988).
Implementation of stochastic quasigradient methods.
In {\it Numerical Techniques for Stochastic Optimization} (eds. Ermoliev, Yu. and Wets, R. J-B), 313--352.
NY: Springer-Verlag.

\bibitem[Hess {\it et al.}(2006)]{Hess2006}
Hess, S., Train, K. E. amd Polak, J. W. (2006).
On the use of a modified Latin hypercube sampling (MLHS) method in the estimation of a mixed logit model for vehicle choice.
{\it Transportation Research Part B}, 40, 147--163.

\bibitem[Hoffman {\it et al.}(2010)]{Hoffman2010}
Hoffman, M. D., Blei, D. M. and Bach, F. (2010).
Online learning for latent Dirichlet allocation.
In {\it Advances in Neural Information Processing Systems 23} (eds. Lafferty, J., Williams, C., Shawe-Taylor, J., Zemel, R. and Culotta, A.), 856--864.
NY USA: Curran Associates, Inc.

\bibitem[Hoffman {\it et al.}(2013)]{Hoffman2013}
Hoffman, M. D., Blei, D. M., Wang, C. and Paisley, J. (2013).
Stochastic variational inference.
{\it Journal of Machine Learning Research}, 14, 1303--1347.

\bibitem[Huang and Wand(2013)]{Huang2013}
Huang, A. and Wand, M. P. (2013).
Simple marginally noninformative prior distributions for covariance matrices.
{\it Bayesian Analysis}, 8, 439--452.

\bibitem[Hubert and Train(2001)]{Hubert2001}
Hubert, J. and Train, K. (2001).
On the similarity of classical and Bayesian estimates of individual mean pathworths.
{\it Marketing Letters}, 12, 259--269. 

\bibitem[Jank(2006)]{Jank2006}
Jank, W. (2006).
Implementing and diagnosing the stochastic approximation EM algorithm.
{\it Journal of Computational and Graphical Statistics}, 15, 803--829.

\bibitem[Jordan {\it et al.}(1999)]{Jordan1999}
Jordan, M. I., Ghahramani, Z., Jaakkola, T. S. and Saul, L. K. (1999).
An introduction to variational methods for graphical models. 
{\it Machine Learning}, 37, 183--233.

\bibitem[Kim {\it et al.}(1995)]{Kim1995}
Kim, B-D., Blattberg, R. C. and Rossi, P. E. (1995).
Modeling the distribution of price sensitivity and implications for optimal retail pricing.
{\it Journal of Business and Economics Statistics}, 13, 291--303.

\bibitem[Knowles and Minka(2011)]{Knowles2011}
Knowles, D. A. and Minka, T. P. (2011).
Non-conjugate variational message passing for multinomial and binary regression. 
In {\it Advances in Neural Information Processing Systems 24} (eds. Shawe-Taylor, J., Zemel, R. S., Bartlett, P. L., Pereira, F. and Weinberger, K.Q.), 1701--1709.
NY USA: Curran Associates, Inc.

\bibitem[Korattikara {\it et al.}(2011)]{Korattikara2011}
Korattikara, A., Boyles, L., Welling, M., Kim, J, and Park, H. (2011).
Statistical Optimization of non-negative matrix factorization.
In {\it JMLR: Workshop and Conference Proceedings}, 15, 128--136.

\bibitem[Lancsar and Louviere(2008)]{Lancsar2008}
Lancsar, E. and Louviere, J. (2008).
Conducting discrete choice experiments to inform healthcare decision making. 
{\it Pharmacoeconomics}, 26, 661--677.

\bibitem[Levin {\it et al.}(2009)]{Levin2009}
Levin, D. A., Peres, Y. and Wilmer, E. L. (2009).
{\it Markov Chains and Mixing Times}
Providence, Rhode Island: American Mathematical Society.

\bibitem[McFadden(1980)]{McFadden1980}
McFadden, D. (1980). 
Econometric models for probabilistic choice among products.
{\it Journal of Business}, 53, 13--29.

\bibitem[McFadden and Train(2000)]{McFadden2000}
McFadden, D. and Train, K. (2000).
Mixed MNL models for discrete response. 
{\it Journal of Applied Econometrics}, 15, 447--470.

\bibitem[Minka(2001)]{Minka2001}
Minka, T. P. (2001).
{\it A family of algorithms for approximate Bayesian inference}.
Ph.D. thesis, MIT.

\bibitem[Opper and Archambeau(2009)]{Opper2009}
Opper, M. and Archambeau, C. (2009).
The variational Gaussian approximation revisited.
{\it Neural computation}, 21, 786--792.

\bibitem[Ormerod and Wand(2010)]{Ormerod2010}
Ormerod, J. T. and Wand, M. P. (2010).
Explaining variational approximations.
{\it The American Statistician}, 64, 140--153. 

\bibitem[Orr(1996)]{Orr1996}
Orr, G. B. (1996).
Removing noise in on-line search using adaptive batch sizes.
In (eds. Mozer, M. C., Jordan, M. I. and Petsche, T. ) {\it Advances in Neural Information Processing Systems 9}, 232--238.
Cambridge MA: MIT Press.

\bibitem[Paisley {\it et al.}(2012)]{Paisley2012}
Paisley, J., Blei, D. M. and Jordan, M. I. (2012).
Variational Bayesian inference with stochastic search.
In (eds. Langford, J. and Pineau, J.) {\it Proceedings of the 29th International Conference on Machine Learning}, 1367--1374.
Omnipress, NY USA.

\bibitem[Powell(2011)]{Powell2011}
Powell, W. B. (2011).
{\it Approximate dynamic programming: solving the curses of dimensionality}, 2nd ed. 
Hoboken, NJ: John Wiley \& Sons, Inc.

\bibitem[Ranganath {\it et al.}(2013)]{Ranganath2013}
Ranganath, R., Wang, C., Blei, D. M. and Xing, E. P. (2013).
An adaptive learning rate for stochastic variational inference.
In {\it JMLR: Workshop and Conference Proceedings}, 28, 298--306.

\bibitem[Robbins and Monro(1951)]{Robbins1951}
Robbins, H. and Monro, S. (1951).
A stochastic approximation method. 
{\it The Annals of Mathematical Statistics}, 22, 400--407. 

\bibitem[Rossi {\it et al.}(2005)]{Rossi2005}
Rossi, P. E., Allenby, G. M. and McCulloch, R. (2005).
{\it Bayesian Statistics and Marketing}.
NJ USA: John Wiley \& Sons.

\bibitem[Salimans and Knowles(2013)]{Salimans2013}
Salimans, T. and Knowles, D. A. (2013)
Fixed-form variational posterior approximation through stochastic linear regression.
{\it Bayesian Analysis}, 4, 837--882.

\bibitem[Spall(2003)]{Spall2003}
Spall, J. C. (2003)
Introduction to stochastic search and optimization: estimation, simulation and control.  
New Jersey; Wiley.

\bibitem[Tan and Nott(2013)]{Tan2013}
Tan, L. S. L. and Nott, D. J. (2013)
Variational inference for generalized linear mixed models using partially non-centered parametrizations. 
{\it Statistical Science}, 28, 168--188.

\bibitem[Tan and Nott(2014)]{Tan2014}
Tan, L. S. L. and Nott, D. J. (2014)
A stochastic variational framework for fitting and diagnosing generalized linear mixed models.
{\it Bayesian Analysis}, to appear.

\bibitem[Train(2009)]{Train2009}
Train, K. E. (2009)
{\it Discrete Choice Methods with Simulation}, 2nd ed.
NY: Cambridge University Press.

\bibitem[Trower(2002)]{Trower2002}
Trower, C. A. (2002)
Can colleges competitively recruit faculty without the prospect of tenure?
In {\it The Questions of Tenure} (ed. R. P. Chait), 182--220.
Cambridge, MA: Harvard University Press.

\bibitem[Wand(2013)]{Wand2013}
Wand, M. P. (2013)
Fully simplified multivariate normal updates in non-conjugate variational message passing.
Available at \url{http://www.uow.edu.au/~mwand/fsupap.pdf}.

\bibitem[Wang {\it et al.}(2011)]{Wang2011}
Wang, C., Paisley, J. and Blei, D. M. (2011)
Online variational inference for the hierarchical Dirichlet process.
In {\it JMLR: Workshop and Conference Proceedings}, 15, 752--760.

\bibitem[Wang and Blei(2013)]{Wang2013}
Wang, C. and Blei, D. M. (2013)
Variational inference in nonconjugate models.
{\it Journal of Machine Learning Research}, 14, 1005--1031.

\bibitem[Waterhouse {\it et al.}(1996)]{Waterhouse1996}
Waterhouse, S., Mackay, D. and Robinson, T. (1996).
Bayesian methods for mixtures of experts.
In {\it Advances in Neural Information Procesing Systems 8} (eds. Touretzky, D. S., Mozer, M. C. and Hasselmo, M. E.), 351--357.
USA: MIT Press.


\end{thebibliography}
\end{document}